# WHITE PAPER ON BUSINESS OF 6G

6G Research Visions, No. 3

June 2020

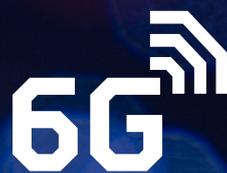

# Table of Contents

**List of contributors:**

**Editors:**
Seppo Yrjölä, University of Oulu, Nokia Enterprise, Oulu, Finland, seppo.yrjola@oulu.fi @nokia.com · Petri Ahokangas, University of Oulu, Oulu Business School, Oulu, Finland · Marja Matinmikko-Blue, University of Oulu, Centre for Wireless Communications, Oulu, Finland

**Contributors in alphabetical order:**
Risto Jurva, University of Oulu, Centre for Wireless Communications, Oulu, Finland · Vivek Kant, Indian Institute of Technology Bombay, IDC School of Design, Mumbai, India · Pasi Karppinen, University of Oulu, OASIS Research Unit, Oulu, Finland · Marianne Kinnula, University of Oulu, INTERACT Research Unit, Oulu, Finland; Harilaos Koumaras, National Center for Scientific Research Demokritos, Institute of Informatics and Telecommunications, Athens, Greece · Mika Rantakokko, University of Oulu, Centre for Wireless Communications, Oulu, Finland · Volker Ziegler, Nokia Bell Labs, München, Germany · Abhishek Thakur, Institute for Development and Research in Banking Technology, Hyderabad, India · Hans-Jürgen Zepernick, Blekinge Institute of Technology, Karlskrona, Sweden

**Please cite:**
Yrjölä, S., Ahokangas, P., & Matinmikko-Blue, M. (Eds.). (2020). *White Paper on Business of 6G* [White paper]. (6G Research Visions, No. 3). University of Oulu. *http://urn.fi/urn:isbn:9789526226767*

6G Flagship, University of Oulu, Finland
June 2020

## Acknowledgement

This white paper has been written by an international expert group, led by the Finnish 6G Flagship program (6gflagship.com) at the University of Oulu, within a series of twelve 6G white papers.

# Abstract

6G vision and the industry consensus of underlying technology enablers have come a long way and will shape the new access, networking, and service domains in future mobile communications. These novel features promise countless opportunities for service innovation and business efficiencies, creating an unprecedented impact on multiple vertical sectors. 6G will connect worlds in novel and innovative ways – the physical and digital worlds will be deeply intertwined in real time, human biological systems will be seamlessly coupled, and at the same time, there will be a new human sensory and cognitive dimension across the scenarios of the 6G experience. Key technology-enabling themes to be explored will include the pervasive leverage of machine learning and artificial intelligence across architectural domains to flexibly define the air interface, as well as service management and orchestration in the 6G "network of networks" topology and platform ecosystem. Terahertz (THz) research is one of the prominent topics, utilizing spectral bands of above 100 GHz for both communications and sensing purposes, thereby enabling connectivity data speeds in the Terabit/s range. We foresee millions of sub-networks and devices becoming the network in conjunction with extreme performance attributes in terms of both sub-millisecond latency, high reliability and time-sensitive determinism, and advanced ways to assure security, privacy, and trust.

In line with 6G vision and technology enablers, developing products, services, and vertical applications for the future digitized society in the 6G era requires a multidisciplinary approach and a redefinition of how we create, deliver, and consume network resources, data, and services for both communications and sensing purposes. This development will change and disrupt the traditional business models and ecosystem roles of digital service providers, as well as open the market for key stakeholders in the 6G era like digital service operators, cloud operators, and resource brokers. Furthermore, sustainable development is a highly complex area that will call for major changes in industrialized society in the long run. This white paper discusses the unprecedented opportunities to enable and empower multiple stakeholders to more actively participate in the future 6G ecosystem via novel sustainable open ecosystemic business models with flexible integration of long tail services with tailored performance attributes. This research adopts a qualitative scenario planning method, portraying three scenario themes resulting in a total of 12 scenarios for the futures of the 6G business. We present both optimistic and pessimistic scenarios, and assess their probability, plausibility, and preferability. By focusing on key trends, their interactions, and irreducible uncertainties, scenario building generates perspectives for the futures within which alternative 6G business strategies have been developed and assessed for a traditional incumbent mobile network operator, and a novel 6G digital service provider stemming from redefined sustainable economics. Value capture in the 6G era requires an understanding of the dynamics of platforms and ecosystems. The results indicate that to reach some of the preferred futures, we should attend to the privacy and security issues related to business and regulation needs: public/governmental, corporate, community, and user perspectives on and aims of governance; ecosystem configuration related to users, decentralized business models, and platforms; user empowerment; and the role of service location-specificity.

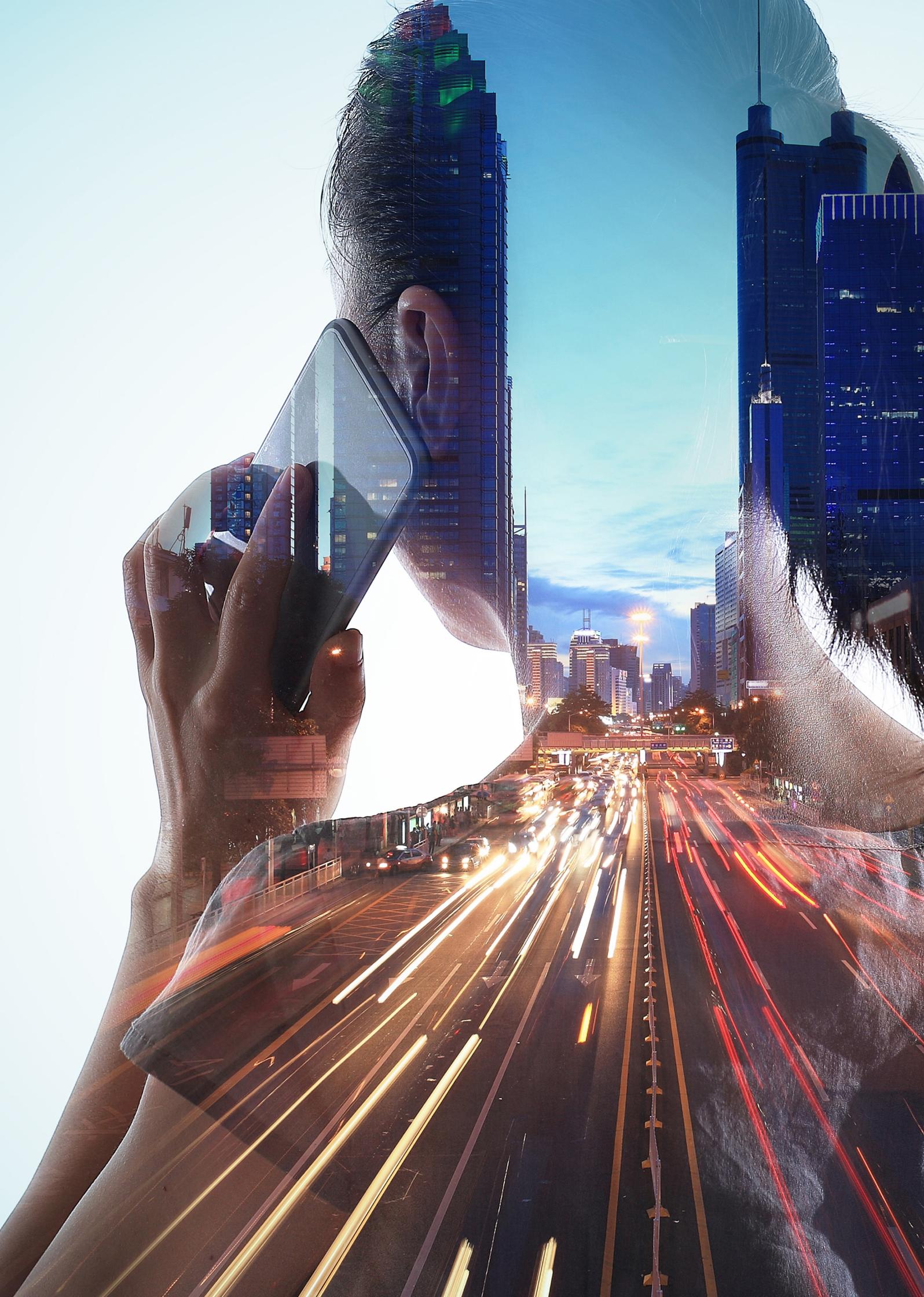

# 1

# Introduction

Novel digital era business models have been transforming and disrupting traditional industries with unprecedented speed, and the telecommunications industry is no exception. The wireless network technology evolution and revolution will transform industries through wireless services provided at speeds of hundreds of gigabits per second, sub-millisecond latencies, support of a wide range of novel applications connecting billions of devices and objects, and versatility by virtualization, enabling innovative business models across multiple sectors (Ahokangas et al. 2018; 2019). The present 5G connectivity market continues to be characterized by incumbent mobile network operators (MNOs), whose business is structured around service mass provisioning with high advance investments in infrastructure and exclusive regulator-granted long-term licenses. At the same time, the responsibility for delivering resources is being transformed from an MNO-centric system into a dynamic mode of operation (Alvarez et al. 2019). This development is due to the 5G deployment of software-defined networks (SDN), Network Function Virtualization (NFV) and network slicing, cloudification, the diffusion of novel local micro operator edge service business models, and the development of vertical service and application ecosystems (Yrjölä et al. 2018).

Futurists, management gurus, trade pundits and general Internet intelligentsia have emphasized that industry 4.0, driven by networking and connectivity, will have a disruptive impact on society. In this line of thought, 6G technologies are ideally positioned to have a massive impact on the communications sector. However, from the perspective of alternative futures research, we must envision 6G technologies holistically from the perspective of interaction between technology and society. In the past decade, there has been a growing consensus about the convergence between nano-, bio-, Infor-, and cogno-based technologies (NBIC convergence). 6G technologies will no doubt ride on this convergence but add to the fabric of society in a very fundamental manner. In turn, this will result in newer forms of business engagement, opportunities and strategies for micro-entrepreneurs, MSMEs, and large conglomerates. From the standpoint of business, a marked change 6G will bring about will be in terms of a greater opportunity in different vertical sectors to generate a considerable revenue pool. The application of big data, artificial intelligence, and cloud computing at the edge of the network with ubiquitous near-real-time wireless connectivity will change many aspects of our personal and working lives, and the structure of the economy. In particular, the diffusion of information and communication technologies in physical industries is poised to increase stagnated productivity growth, especially from close integration with digital twins.

Furthermore, the future of 6G telecommunications will be shaped by growing societal requirements like inclusivity, sustainability, resilience, and transparency – a highly complex area that will call for major changes in industrialized societies in the long run (Latva-aho & Leppänen 2019). As the basic connectivity service continues to be commoditized as essential for any standard of living, the telecommunication industry is exploring new ways to better position itself for the digital transformation and going beyond the traditional role of connectivity provisioning (Ahokangas et al. 2019). Somewhat surprisingly, compared to other infrastructures like energy or water supply, commoditization is happening only after five decades of mobile communication infrastructure deployment. Access to data and data ownership are increasingly the major factors in value creation, and limiting such access is a means of control and restricting empowerment. Creating a 6G system that transforms how data is collected, shared, analyzed, and acted on in real time may create strong drivers for future value and introduce novel stakeholder roles, but it may also lead to serious privacy and ethical concerns about the location and use of data. The pervasive influence of artificial intelligence will not merely reflect what something looks like but its context, meaning, and function, creating an Internet of Skills, Internet of Senses, and digital twins, while enabling trust and assuring security and privacy (Latva-aho & Leppänen 2019).

The preceding discussion is indicative of the increased importance of novel business model thinking in 6G from a triad of design, engineering, and economics perspectives. Most successful technologies will be at the intersection of this triad. Design practice is of crucial importance to 6G technologies, because it will be the tipping point for many businesses in the 6G ecosystem. Design fuels the innovation, consumption, and development of newer forms of interaction that will have a crucial impact on developing a supply focus mentality. Such a mentality will enable the creating, leveraging, and sustaining value for the customer and society as a whole. With the rapid convergence of media technologies—radio, television, print, computers, Internet, mobile—in recent decades, there has been a shift in the public consumption of media and interaction with technologies. Marshall McLuhan's dictum of "the medium is the message" is rapidly coming to the fore. Governments, businesses, and users alike are clamoring for the human-centered design of technologies. As a result, we may be observing a rise of a new demographic of 6G consumers, whom we label "experientials." These experientials are similar to other target demographic labels used by businesses to make sense of their potential consumers, such as "millennials" and "netizens," among others. However, experientials differ from other labels because of their central focus on the appropriate user experience. They may come from different walks of life and occupations such as factory workers or homemakers, but they are all joined around the user experience as the central idea of interacting with 6G technologies. These demographics will in turn have a massive impact on the manner in which design, engineering, and economics must be addressed coherently, opportunistically, and strategically.

Engineering research, stemming from product and manufacturing platforms, and lately service modularity (de Mattos, Fettermann & Cauchick-Miguel 2019), focuses on components and interfaces that aim to create economies of scale. In parallel, economics research discusses how to connect demand and supply to grow sustainably and enter or create new markets (Gawer 2014). What both streams agree on is that platforms create an ecosystem around them, paving the way to seeing platforms and ecosystems as intertwined (Teece 2018). Furthermore, a recent study (Yrjölä, Ahokangas, & Matinmikko-Blue 2019) discusses how the transformation from current network-for-connectivity business models to a network-of-services model builds on the platform with data and algorithms.

Existing 5G business studies focus on traditional MNO business models and discuss 5G in rather technical and general terms, mostly at the industry level (Koumaras et al. 2018), while platform or ecosystemic business models have seldom been examined (Ahokangas et al. 2018; Yrjölä et al. 2018). Collaborative business models have been introduced (Noll & Chowdhury 2011), as well as related system integrator, neutral host, and brokerage roles (Ballon 2009; Rasheed et al. 2016; Valtanen, Backman, & Yrjola 2019). Operators' capabilities of exposing network functionalities through adopting web-based service models are analyzed (Gonçalves & Ballon 2011), as is the utilization of the cloud business model (Zhang et al. 2015). Moreover, the localized nature of 5G services has emerged as a characteristic in these studies (Ahokangas et al. 2016) and introduced the micro-operator concept (Matinmikko et al. 2017). Beyond technicalities, the discussed business models can be seen as representing two basic mobile operator business models, including the connectivity service provider and its differentiation (Ahokangas et al. 2013, 2016). As an emerging field, 6G business models have not been widely discussed in the literature to date. However, vision papers on future communication needs, enabling technologies, the role of AI, and emerging use cases and applications have recently been published (Viswanathan & Mogensen 2020; Latva-aho & Leppänen 2019; Saad, Bennis, & Chen 2019; Letaief et al. 2019; Katz, Pirinen, & Posti 2019). Discussion has recently expanded to 6G indicators of value and performance (Ziegler and Yrjölä 2020), the role of regulation and spectrum sharing (Matinmikko-Blue et al. 2020), and the antecedents of multi-sided transactional platforms (Yrjölä et al. 2020).

Building on the above discussion, with roots in engineering and economics research, this white paper seeks to develop alternative scenarios for 6G business over 10 to 15 years (by 2030–2035), and derive the key trends and uncertainties shaping the future. Furthermore, this paper's theoretical research question is: How can the 6G business be transformed from closed connectivity-driven business to novel open sustainable ecosystemic business models in 6G? This research follows the anticipatory action learning method (Stevenson 2002; Inayatullah 1998, 2005) and adopts the scenario planning process (Schoemaker 1995; Schwartz 1991). The data utilized in this paper is based on a set of virtual future-oriented white paper expert group workshops organized by the 6G Wireless Summit 2020 (6G Summit 2020). The paper is structured as follows: Section 2 describes the research methods and theory frameworks adopted; Section 3 discusses identified trends and uncertainties, and presents developed scenarios; Section 4 analyzes the implications of the scenarios and their strategic options; Section 5 draws conclusions and highlights perspectives for future research.

# Methods and theory frameworks

**2**

## 2.1 Scenario planning process

Exploratory scenarios describe events, trends, and choices as they might evolve based on alternative assumptions as to how these events, trends, and choices may influence the future. Exploratory scenarios provide a plurality of plausible alternative futures, and they can be created through the *anticipatory action learning process* (Stevenson 2002; Inayatullah 1998, 2005) in which professionals from different fields come together to generate scenarios. The alternative futures presented in this white paper were created and assessed through groupwork, and special attention was paid to the coherence, variation, and validity of scenarios (Stewart 2007; Collins & Hines 2010). The data for the alternative 6G business futures presented in this paper were mapped in the 6G Flagship program at the University of Oulu, funded by the Academy of Finland. The scenarios were generated by professionals in facilitated scenario workshops during the first four months of 2020. The scenario work process discussed in this chapter was carried out in teleconference sessions and individual homework consisting of the following steps (Schoemaker 1995; Schwartz 1991) depicted in Figure 1:

· Selecting the focal issue and time frame: Business of 6G 2030–2035
· Discovery and identification of key factors and driving forces
· Identification and evaluation of forces by choosing key trends and uncertainties based on their anticipated impact and predictability of consequences and uncertainty
· Establishing the scenario logic representing the most significant uncertainties of the overall system under scrutiny by selecting two unrelated polar dimensions
· Creating the scenarios by building on different perspectives and the identification of trends and uncertainties, and by crossing their outcomes in a scenario matrix
· Assessment and evaluation of different scenarios based on their internal consistency, depth, richness, plausibility, and key stakeholders' behavior (Voros 2003)
· Broadening and deepening scenario futures thinking, utilizing Causal Layered Analysis (Inayatullah 1998) and Integral Theory four perspective models (Stewart 2007).

A total of 146 forces was identified, and 12 scenarios were generated by the workshop participants. The workshop composition ensured a wide variation in participants' backgrounds, which included research, standardization and development, academia, the telecoms industry, government, and verticals. The created scenarios were *backcasted* (Dreborg 1996) to technology-themed 6G White Paper workshop results to analyze key strategic technology options and frames against scenarios. This enabled the outlining of research targets and questions, and the selection of indicators and signposts to monitor progress. Backcasting entails defining a desirable future and then working backwards to identify policies and programs that will connect it to the present. Foresight scenarios are by definition future-focused, and their reliability and validity cannot be directly controlled. Instead, the quality of research can be evaluated by how *probable*, *plausible*, and *preferable* the outcome appears to be. These views have been incorporated in the scenario planning process in Section 3.

The implications of the developed scenarios for future businesses were discussed utilizing the *simple rules strategy framework* (Eisenhardt & Sull 2001; McGrath 2010). The simple rules framework consists of six themes as rules: the nature of *opportunity* rule; *how-to* rules for conducting business; *boundary* rules for defining the boundaries of the business; *priority* rules, which help to identify and rank decision-making criteria; *timing* rules, which help in identifying, sequencing, synchronizing, and pacing things; and *exit* rules, which help in identifying the basis for exit or selecting things to be stopped/ended or given up. This viewpoint for future 6G business is developed further in Section 4.

## 2.2 Business model concept and value configuration

The business model concept centering the value creation processes has become the tool for making boundary-spanning analyses in contemporary business research (Zott, Amit, & Massa 2011). Traditional business model definitions from the activity perspective (Onetti et al. 2012) assume a focal firm: *"We define the business model as the way a company structures its own activities in determining the focus, locus and modus of its business,"* while more recent views, e.g. Amit & Zott (2001), do not necessitate a focal point: *"a business model depicts the design of transaction content, structure, and governance so as to create value through the exploitation of business opportunities."* Furthermore, business models are seen to connect to three strategic choices and related key activities of firms within ecosystems: 1) *opportunities* to be explored and exploited; 2) the *value* to be created and captured; and 3) *advantages* to be explored and exploited (Morris, Schindehutte, & Allen 2005; Teece 2010; McGrath 2010; Zott, Amit, & Massa 2011). Exploring and exploiting opportunities and advantages can be seen to motivate ecosystemic interaction from a dynamic capability perspective (Gomes et al. 2018), whereas value creation, delivery, sharing, and capture are considered the key elements of a functioning business model (Osterwalder & Pigneur 2010). Similarly, successful business models are considered to have three strategic consequences: 1) *scalability* (Stampfl, Prügl, & Osterloh 2013); 2) *replicability* (Aspara, Hietanen, & Tikkanen 2010); and 3) *sustainability* (Schaltegger, Hansen, & Lüdeke-Freund 2016). The growth of business is frequently connected to scalability and replicability. While scalability refers to internal growth potential and flexibility, replicability indicates its external flexibility to adapt. Moreover, sustainability stems from a business model's feasibility, viability, and environmental or societal impact.

In its scenario analysis, this white paper utilizes the integrated business model configurations and value configurations framework (Xu 2019). The classical business model conceptualization and value discussions build on Porter's (1996) value chain theory and a *supply-focused* mentality that sees the business model as a means to capture value from customers (Massa, Tucci, & Afuah 2017), and where the producer with the system of activities and resources is the sole creator of the value, a focal firm (Casadesus-Masanell & Ricart 2011). In contrast with classical approaches, the *demand-focused* view shifts from the supply-focused business model configuration, emphasizing the utilization of customer interaction mechanisms to enable value co-creation incorporating a means of creating and delivering value for a target group of consumers, and a means of using existing resources and processes to promote the stable interaction of mechanisms (Bereznoi 2015). This white paper adopts a service-dominant logic and service ecosystem thinking to analyze developed scenarios and the potential for the transformation of the wireless industry from traditional supply-focused regimes to ecosystem-focused models.

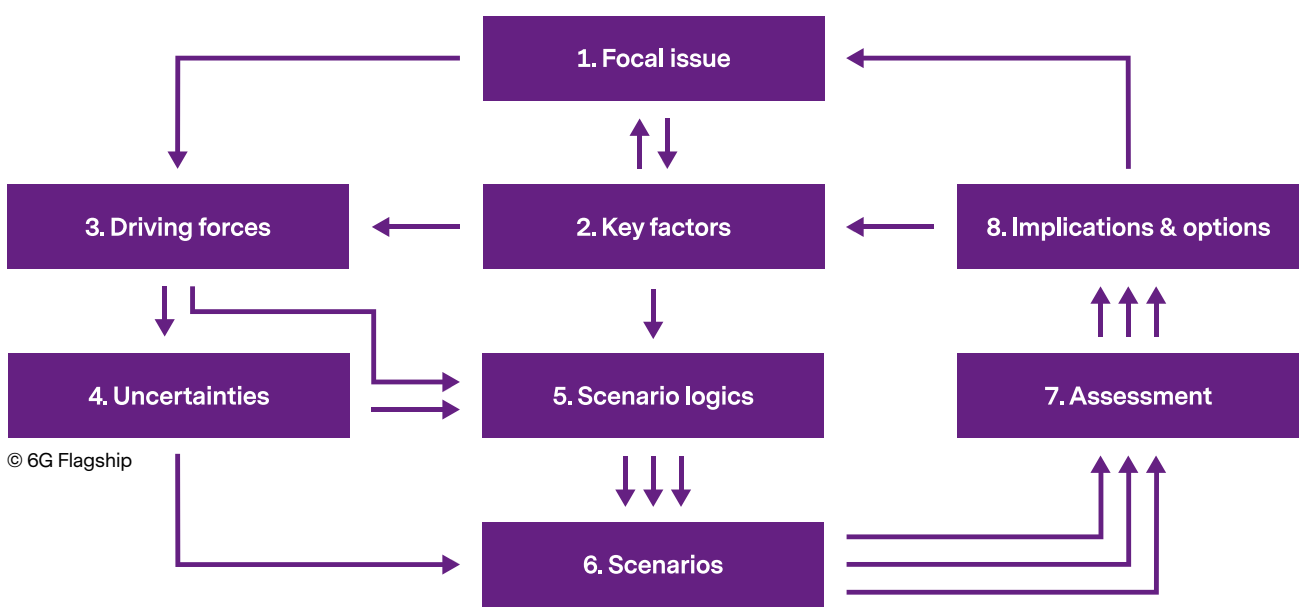

© 6G Flagship

**Figure 1. The scenario planning process**

Vargo and Lusch (2016) define a service ecosystem as a self-adjusting system of resource-integrating actors connected by shared institutional arrangements and mutual value creation through service exchange. This demonstrates the systemic and institutional nature of value co-creation and resource integration: if all actors/stakeholders are to be willing to bring their resources to the common pool, i.e. continue collaboration within the ecosystem, they all need to experience value from it. The value perspective of the *ecosystem-focused* business model is about value co-creation and co-capture to maximize the overall ecosystem value, which in turn increases the value shared and acquired not only by a focal firm, but by the actors within the ecosystem (Xu 2019). From a structural perspective, the ecosystem can be seen as having four aspects: 1) network governance; 2) platform keystone agents and complementors; 3) open interfaces and a pool of innovative capabilities and resources; and 4) a modular core and periphery design (Mazhelis & Mazhelis 2012). These aspects of the ecosystems have shaped our analysis, and they are further discussed in Sections 4 and 5.

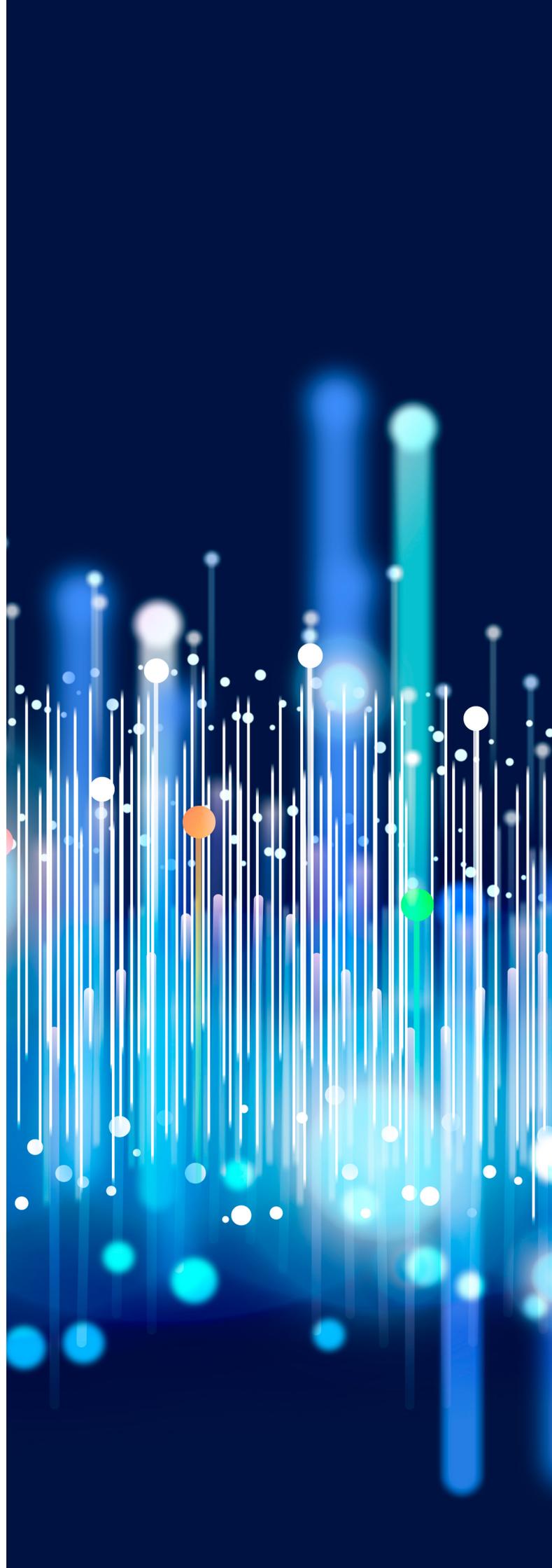

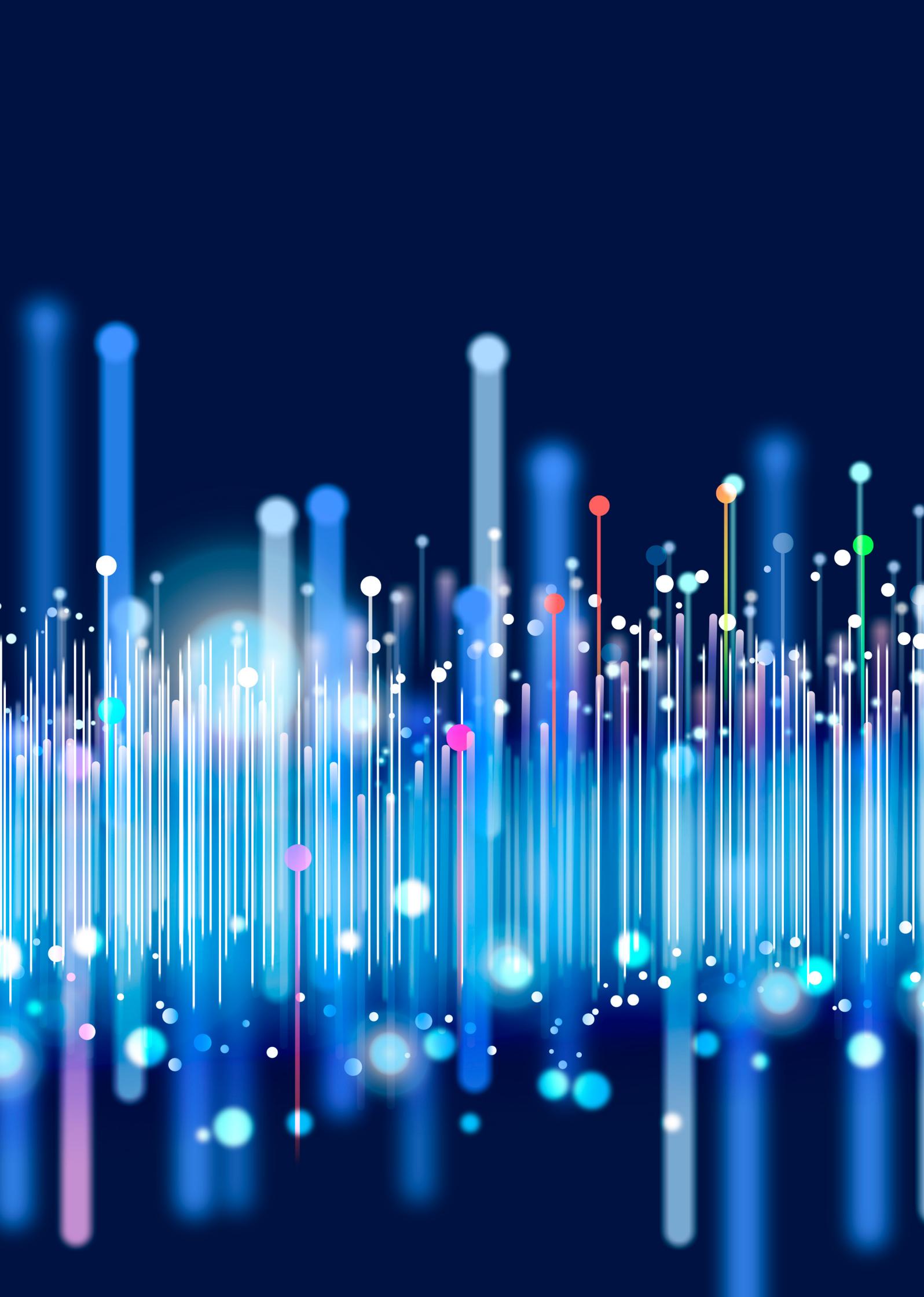

# Developed scenarios for 6G business

**3**

## 3.1 Forces—Trends and uncertainties

To gain an understanding of the drivers of future 6G business, the workshop participants discovered and identified a total of 146 forces. These participants represented research, standardization and development, academia, the telecoms industry, government, and other related verticals. As secondary data, selected recent trend analysis work was utilized (ABI Research 2020; Analysys Mason Research 2020; Business Finland 2020; DTTL 2017; Dufva M 2020; Ericsson 2020; ESPAS 2020; FG-NET-2030 2019; Frost & Sullivan 2020; GSMA 2019, 2020; Huawei 2019; McKinsey Global Institute 2019; Oliver Wyman & Nokia 2020; Ripple W. J et al. 2017; UN 2019; WEF 2020).

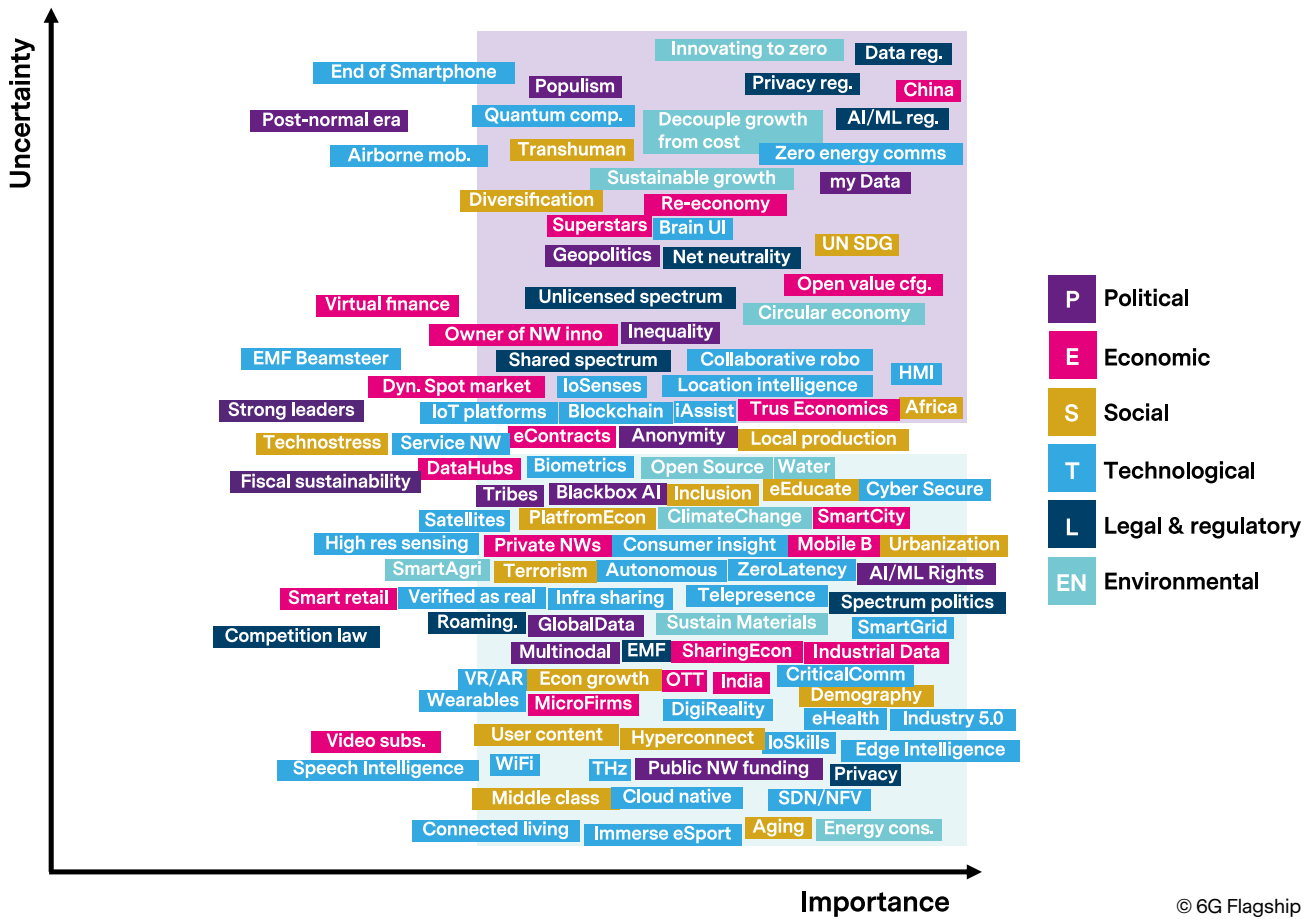

Figure 2. Summary of the evaluation of forces on uncertainty and importance

© 6G Flagship

Key trends were selected from the identified forces by identifying those with the highest impact/importance and the lowest uncertainty discussed. The trends by definition appear in all the developed scenarios. However, key uncertainties were selected from the forces based on the highest impact/importance, highest uncertainty, and highest variances in the participants' ranking. A summary of the evaluation of all considered forces is depicted in Figure 2, and the key trends and uncertainties identified from these forces are shown in Figure 3, using the PESTLE framework.

### Key trends

Based on the ranking of forces, several trends were identified, as shown on the left-hand side of Figure 3. They are further discussed in the following. *Public network funding* has traditionally been directed at unserved and underserved areas in terms of broadband access and coverage. Recently, support for deployment programs has extended to policy areas such as smart city community development, worksites and ecosystems/hubs (such as harbors and airports), advanced health services, logistics and transport, Smart Cities, public safety, and critical infrastructure at length. It is expected that at the current rate of growth, smart grids will be extended to a variety of sectors, including electricity, the Internet, and healthcare in the future. All are expected to be hyperconnected and completely automated. They will serve as a middle layer between humans and natural environments, enhancing the capabilities of both. These networks will be put together with a public-private-personal ownership funding model with a view to sustainable growth and the use of digital infrastructure.

There are contrary interpretations of the *artificial intelligence rights* trend. Assuming the availability of appropriate datasets for training purposes, artificial intelligence can propose solutions to increasingly complex problems that can serve as the source of great economic growth, shared prosperity, and the fulfillment of all human rights. In the alternative future, it will drive inequality, inadvertently divide communities, and will even be actively used to deny human rights.

*Over-the-top (OTT) companies* utilize their customer data, cloud infrastructure and AI/ML capabilities to challenge traditional operators' customer relationship ownership, because users value service experience and perceive connectivity as a basic utility. In addition to the media space, OTT players offer basic telco services such as voice or messaging, and are active in growth areas such as cloud space and services, competing with telcos for clients and revenue. They tie customers to their own ecosystems with carrier-neutral connectivity, while making reliance on traditional operators a thing of the past.

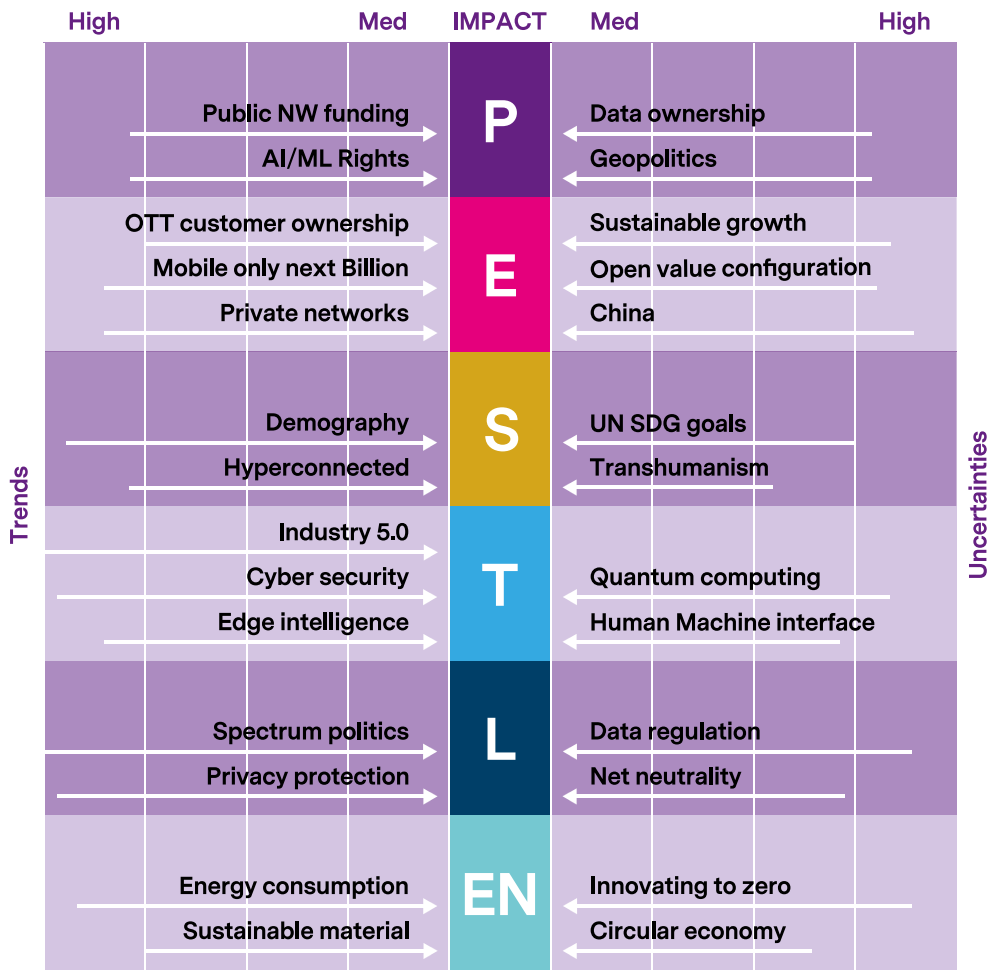

**Figure 3. Identified key trends and uncertainties from the forces**

© 6G Flagship

In *Mobile-only next billion*, ubiquitous cheap phones and increasingly affordable network connections in megacities and rural areas will help another billion users join the Internet and increasingly access applications and digital content aimed at non-English speaking markets. Currently, mobile is for many the primary or only channel for accessing the Internet and services. With its unprecedented scale and growing impact on daily lives, mobile is a powerful tool for achieving the United Nations' Sustainable Development Goals (SDGs) and driving sustainable economic growth.

In 2015, 85 percent of global GDP was generated in cities. *Urbanization* will see 5 billion people living in cities by 2030, occupying 3 percent of the Earth's land, but accounting for 80 percent of energy consumption and 75 percent of carbon emissions. Ninety-five percent of urban expansion in the next decades will occur in the developing world, where 883 million people live in slums today. Rapid urbanization is exerting pressure on fresh water supplies, sewage, the living environment, and public health. Future *demographics* divide a world of 8.6 billion inhabitants by 2030 into two camps: a growing one in Sub-Saharan Africa and South Asia; and a stalling and decreasing one, including Europe, Russia, and post-peak China. Future cities will be hungry global economic engines and the economic powerhouses of the global economy. Cities are increasingly functioning as autonomous entities, setting social and economic standards. Urban identity will grow in importance compared with national identity.

*Private networks* driven by industrial digital automation call for standalone networks for high reliability, high performance in terms of both bandwidth and reliability, secure communications, and data privacy, fulfilling business and mission-critical needs. The solutions will enable the integration of processes, data, and diverse devises such as sensors, machines, and in-vehicle and hand-held devices across a wide range of applications for industry enterprises. Private networks can be established without direct MNO involvement. Furthermore, demands for privacy in personal space may also create private networks that rarely connect with the public Internet.

A *hyper-connected* globe will continue to feel ever smaller in 2030: Globally, 90 percent will be able to read, access the Internet, and be on the move. The aim is to recognize that 6G will transform urban and rural living, existing at the intersection of geopolitics, the growth of nationalism, rights to information transparency, and information citizenry. Thus, once the 6G infrastructure is in place, content growth will lie in supporting multiple social and technological identities of people through a variety of media. This will require a mindful view of decision making and the regulation of future data, information, media, and network usage in the light of sustainable economic growth. Thus, the human in 6G worlds will be increasingly sophisticated in their media and service consumption, while being rooted in their local economies. Connectivity will therefore be not only virtual and digital, but physical. Furthermore, the approaching opportunity to redefine the human–machine interface will enable the biological world to be connected in novel ways.

*Industry 5.0 (I5.0)* will allow collaborative human–machine interaction (HMI) across services and industries, because human intelligence is in perfect harmony with advanced cognitive computing. With real-time data, effective data monetization, and digital automation of the manufacturing process, businesses will be able to shift the focus to generating higher revenues from the *servitization* of products. Advanced manufacturing capabilities will help to overcome design complexities with its ability to facilitate the extremely long tail of mass customization and further return control to customers in a haptic way. Furthermore, I5.0 will require the highest standards of safety and environmental protection.

The need for *cybersecurity and trust* will be ubiquitous in the hyper connected world of 2030. Even a temporary loss of technology may have not only a productivity but a psychological impact on our lives. Furthermore, the subversion or corruption of our technology may result in disastrous harm to our lives and businesses if e.g. medical treatment devices deliver the wrong medication, education systems teach propaganda, or work automation causes injury or damage to our products and businesses. In particular, expectations of protecting and safeguarding society and critical infrastructure from emergency situations by means of technological advances are anticipated to grow.

With the growth of *edge and extreme edge intelligence*, the proliferation of increasingly powerful communication, computing, and analytics resources at the edge of the network will convert architecturally disaggregated 6G access networks into a rich service provisioning and access platform. Hyper-local services such as augmented reality scenarios do not require connectivity with a distant service platform. Instead, they perform better, with local real-time service access. Furthermore, the individual will support the parts of shared information processing and edge intelligence networks that address collective problems for humanity, such as genome sequencing, through shared resources (à la citizen science). The individual will emerge as a node in the network of intelligence relations, rooted in the local physical world while connected to the hyper-real 6G intelligence networks. This will adopt a public good perspective through a 6G digital infrastructure by an ecology of information devices, products, and IoT/IoE services.

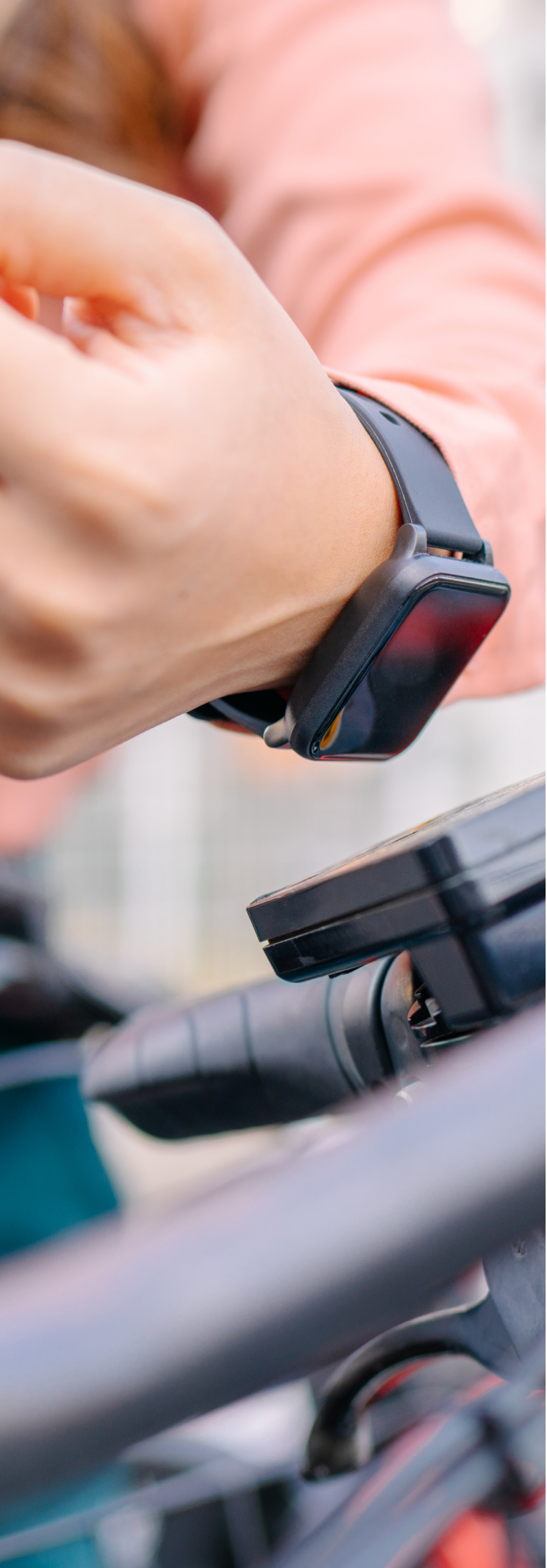

*Wireless spectrum politics* and spectrum management in the 6G era will reveal a new level of complexity that stems from the variety of spectrum bands and spectrum access models with different levels of spectrum sharing. Local deployments of networks by a variety of stakeholders are expected to grow further in 6G. The timescales of international-level spectrum management will no longer be enough with the rapid technology development of mobile communication networks and changing user needs. Spectrum sharing will play an increasing role in accommodating new 6G systems with existing spectrum users. Furthermore, in national technology politics, spectrum regulation is used to gain a competitive advantage (Matinmikko-Blue, Yrjölä, & Ahokangas 2020).

*Privacy regulation* is strongly linked to the rising trends of the platform data economy, p2p sharing economy, intelligent assistants, connected living in smart cities, transhumanism, and digital twins' reality at length. "I own my data" is expected to grow, particularly in Europe, based on GDPR evolution. However, severe differences in global data privacy laws are expected to be living on borrowed time. For example, the US is unique among major countries in lacking a unified set of data privacy laws in spite of having a large number of global web-scale companies, and China's Cybersecurity Law (CSL) applies not only to conventional data handlers but to telecom, radio, and television operators. This is unique, because the Chinese authorities must be informed if data indicates any prohibited activity.

*Sustainable material* will contribute to *Innovating to zero* and *circular economy* megatrends. Toward 2030, companies will shift focus, developing products and technologies that innovate to zero, including zero-waste and zero-emission technologies. The full lifecycle carbon footprint of the ICT industry represents around 2 percent of worldwide emissions and is projected to grow at a 6 percent annual compound growth rate. 6G net positive impact and sustainability are expected to be achieved by enabling increased efficiencies and improved environmental performance in other sectors. Computing technologies will be miniaturized to the extent that they sustain the power generated by everyday human activity. Everyday walking, jogging, and housework will produce the energy to support the person's information devices, which will in turn occasionally monitor the person's vitals, as well as cater to information and entertainment needs through over-the-top connectivity.

## Key uncertainties

As a result of force ranking, key uncertainties were identified, as shown on the right-hand side of Figure 3.

In *geopolitics,* the tension between globalization, networking power, and the urgency of ecological reconstruction will be linked to the balance between centralized decisions and the strengthening of inclusion and democracy. Toward 2030, *power configuration* may be transforming from a multi-polarized world to a poly-nodal world in which power will be determined in economic, technological, and cultural networks and interaction. Political and societal systems face growing tension in responding to the instability of the financial situation, the ecological sustainability crisis, and uncertainty about the complexity of the future world. Societies may struggle to find a balance between fast-moving decision making, community engagement, and the reasserting of democratic values and commitments. On the one hand, it is hoped that strong leaders will bring simplicity to complex problems, but on the other, there could be increasing efforts to influence things in communities from the grassroots level. Furthermore, with increasing polarization, the aging that emerged first in developed countries, and the diversification of the population, new tribes and communities will emerge around various imaginary groups representing a wide variety of values, places of residence, political opinions, consumption choices, or lifestyles. It may happen that weakened and fragmented future prospects, the absence of togetherness, and the polarizing effect of social media will lead to a rise in populism, skepticism toward changes in the environment, and in the worst case, extremist attitudes. At the same time, environmental awareness among people and companies may increase and be reflected in a growing number of people and communities changing their habits, and companies taking corresponding actions to offer customer experience. Vehicles of open value creation and the open source paradigm in particular may provide a powerful avenue to reinvent civil society's participatory process in conjunction with regulatory capability.

*Resource orchestration and configuration* relates to power over development and adoption of innovations and technology that are ubiquitously embedded in society and daily life. Data is increasingly accumulated, and its value and significance are growing. Technology may increasingly be seen as a geopolitical issue of power, and questions of future resource orchestration power will emerge: Who will own the continuously accumulating data? who will get to decide on technology? and who will set the rules and regulations?

*Open value configuration* will emphasize value co-creation and co-capture to maximize the overall ecosystem value, which in turn may increase the value shared and acquired not only by a focal firm but by the actors within the ecosystem. In utilizing sharing and circular economy trends, co-creation employs existing resources and processes to promote the stable interaction of mechanisms. Toward 2030, platform ecosystems will not only offer search, social media, and ecommerce but provide an infrastructure in which innovation and transaction platforms are built. Novel decentralized business models will not necessitate a focal point but depict the design of transaction content, structure, and governance to create value. *Transhumanism* reflects the rise of technology-driven evolution at an unprecedented rate of change, prompting deeper questions into what it is to be human from biological, behavioral, and human-machine evolutionary perspectives. By 2030, we could see a greater societal focus on sustainability, the nature of humanity, values, creativity, and self/social fulfillment and empowerment (Kinnula & Iivari 2019). There may also be a cognitive intelligence revolution via the ascendancy of sentient tools and possibly also a human-orchestrated self-directed selection in biological, neurological, and physical evolution.

Alternative compute approaches such as *Quantum computing* as opposed to classical "calculus" computers is at its best in sorting, finding prime numbers, simulating molecules, and optimization, and could thus disrupt segments like finance, intelligence, drug design and discovery, utilities, polymer design, AI and Big Data search, and digital manufacturing. Technology may for a long time be limited to selected industries such as the military, national laboratories, and aerospace agencies, while alternative compute approaches to help handle the greatly increasing level of parallelism in algorithms may be available more widely.

According to *Net neutrality*, ruling Internet access providers should treat all traffic equally, irrespective of sender, receiver, content, service, application, or the device in use. At the same time, the 5G evolution is already developing a network that can be extremely tailored to specific use cases intending to treat traffic differently for each use case. This legislation creates uncertainties by impacting companies' capabilities to create and capture value in virtualized network-based services between telecom operators and cloud providers. One of the key uncertainties concerns how edge computing should be provided under strict net neutrality, e.g. as in Europe. Furthermore, it impacts capabilities of providing the cyber security required for the merging of the physical and digital worlds that is now happening (Kantola 2019).

## 3.2 Novel 6G business scenarios

Based on the collected data and identified trends and uncertainties, the following six scenario logic dimensions and their endpoints were selected to develop future 6G business scenarios, as summarized in Figure 4. The identification of forces was done by choosing key trends and uncertainties based on their anticipated impact and the predictability of consequences and uncertainty. The scenario logic was selected to represent the most significant uncertainties of the overall system under scrutiny by selecting two unrelated polar dimensions. The six scenario dimensions were categorized into three themes to develop a total of 12 alternative 6G business scenarios, which are discussed next.

### 3.2.1 Scenario theme #1 – User experience

In the User experience scenario theme, we chose the horizontal dimension to present resource orchestration and the vertical axis user experience, as depicted in Figure 5. The polar dimensions opposite to the resource orchestration axis are societal/corporate and individual-driven orchestration. User experience logic ends are traditional standardized service provisioning and the opposite driving customer engagement with customized long tail service experiences. Using these two scenario dimensions, we have developed four scenarios, as shown in Figure 5 and discussed in the following.

#### Customer6.0 (1A) Customized experience and resource orchestration by user

In scenario Customer6.0, user experience is customized, and resource orchestration is user-centric. 6G technology has penetrated most parts of the world. IoT devices and sensors controlled by AI are a normal part of the environment nearly everywhere. Automatic collection of different kinds of data from humans, as well as from our environment and its analysis, are used for highly sophis-ticated products and systems that make people's lives easier and provide better user experience through convenience, because everything is automated. The prices of the systems are very reasonable due to opened-up interfaces and standardized cheap components. Computing technologies are miniaturized to the extent that they sustain the power generated through everyday human activity to support the individual's information devices, which in turn occasionally monitor the person's vitals, as well as catering to information and entertainment needs through over-the-top connectivity. Media and service consumption are rooted in local economies, and users of such products and systems are used to living with them and cannot imagine their lives without them.

Hyperconnected and completely automated networks have been put together with a public-private-personal ownership funding model with a view to sustainable growth and digital infrastructure usage. As counterforces to the creation of platform monopolies, decentralized platform cooperatives, the peer-to-peer economy, shared economy models, and the progress of a human-driven fair data economy have emerged. Transhumanism reflects the rise of technology-driven evolution at an unprecedented rate of change, prompting deeper questions about what it is to be human from biological, behavioral, and human-machine evolutionary perspectives. By 2030, we could see a greater societal focus on sustainability, the nature of humanity, values, creativity, self-/social fulfillment, and empowerment. There may be a cognitive intelligence revolution via the ascendancy of the sentient personal assistant and possibly human-orchestrated self-directed selection in biological, neurological, and physical evolution. The emerging opportunity to redefine human–machine and brain–UI interfaces enables the connecting of people and the biological world in novel ways. Holopresence systems can project realistic, full-motion, real-time 3D digital twin images of distant people and objects into a room, along with real-time audio communication, with a level of reality rivaling physi-

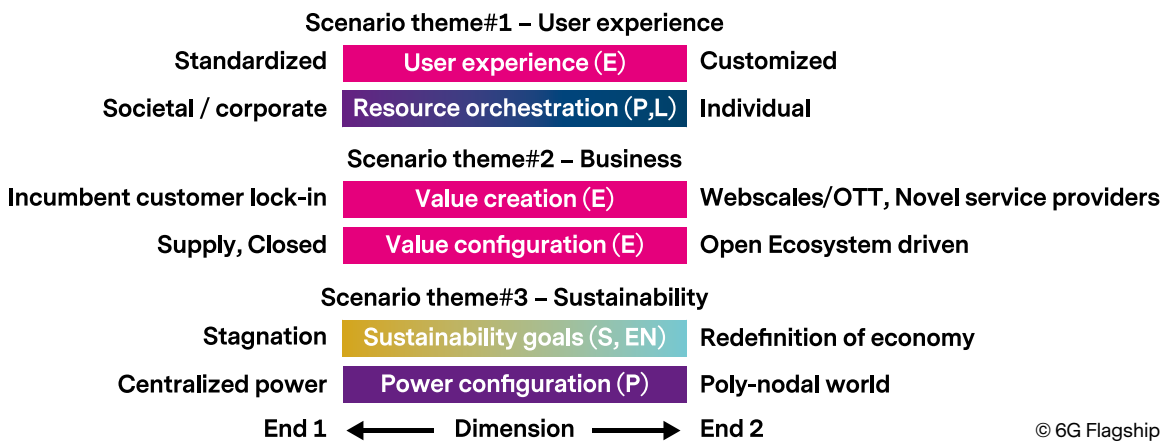

Scenario theme#1 – User experience

| Standardized | User experience (E) | Customized |
| Societal / corporate | Resource orchestration (P,L) | Individual |

Scenario theme#2 – Business

| Incumbent customer lock-in | Value creation (E) | Webscales/OTT, Novel service providers |
| Supply, Closed | Value configuration (E) | Open Ecosystem driven |

Scenario theme#3 – Sustainability

| Stagnation | Sustainability goals (S, EN) | Redefinition of economy |
| Centralized power | Power configuration (P) | Poly-nodal world |

End 1 ◄——— Dimension ———► End 2

© 6G Flagship

**Figure 4. Summary of selected scenario logics**

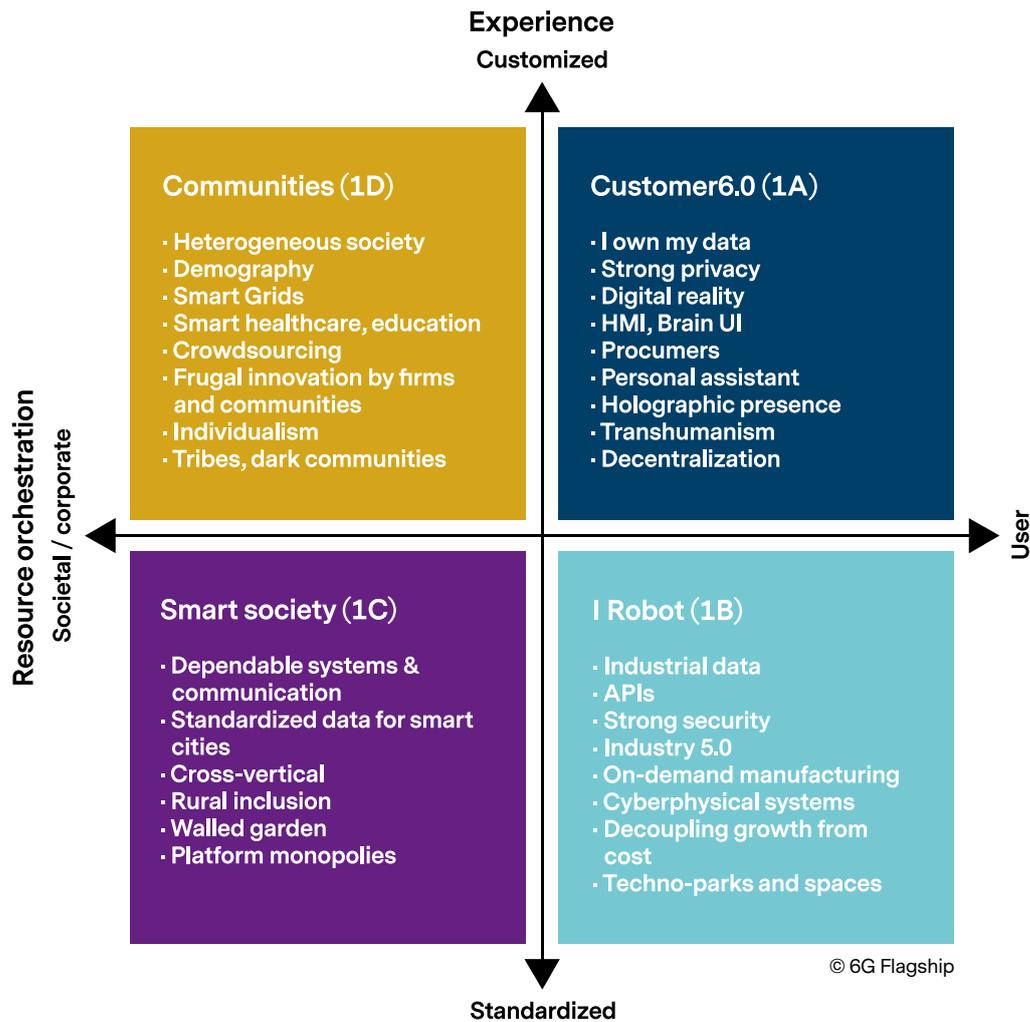

Figure 5. Scenarios based on user experience and resource orchestration

cal presence. Images of remote people and surrounding objects are captured and transmitted over a 6G network and projected using laser beams in real time.

There will be a serious threat of a digital divide and inequality related to access and skills to use new technologies, knowledge, digital services, and materials at the individual level, as well as between countries, if access to new technologies is restricted. This may reflect one's working career, which also assumes employee activity to educate individuals. Expensive products and systems may never be the norm, but both local businesses and citizens can create their own frugal adaptations of products and systems to suit their living conditions when even electricity may be a scarce resource. This is supported by the global do-it-yourself culture, in which the sharing of blueprints and working processes is encouraged. 6G will transform urban and rural living at the intersection of geopolitics, the growth of nationalism, rights to information transparency, and information citizenry. Content growth will lie in supporting people's multiple social and technological identities through a variety of media. This will require a mindful view of decision making and the regulation of future data, information, media, and network usage in light of sustainable growth for the economy and human empowerment.

## I Robot (1B) Standardized experience & resource orchestration by user

In the I Robot scenario, user experience is standardized, and resource orchestration is user-centric. With the convergence of nanotechnology, biotechnology, information technologies, and cognitive science (NBIC), newer application areas, goods, services, and systems will proliferate. This convergence has resulted in our present development of cyber-physical systems and IoT-based technologies, along with 3D printing and on-demand manufacturing, among other instantiations. In the next two decades, the growth of 6G-enabled technologies will aid the explosion of biologically based intelligence along with artificial intelligence in industrial setups. This biological intelligence will rely on a mixture of biologically based self-programmable natural and artificial neural networks and micro- and nanobots that can be used in tandem with existing AI-based automated systems. Interaction with these hybrid bio-industrial automated systems will constitute the next major revolution in programmable smart factories and industrial systems, and it will be a source of value creation, configuration, and resource utilization. In a rapidly changing reprogrammable and reconfigurable world, businesses will have a short to medium change horizon, expected performance indicators, and return on investments. The market of mid-level businesses in the industrial sector will be increasingly facilitated via agile and scalable technoparks and spaces. The people, processes, and resources needed for such mid-scale businesses and services will require flexibility and rapid learning to transfer learning from one job to another. This will enable the overall value optimization to be based on the level of units of businesses housed in one technopark, regardless of their individual sectors.

While countries will continue the movement of goods in a globalized world, the nature of transfer will have a marked impact. In a world enabled by 6G technologies and 3D printing, the blueprint will be delivered to proprietary machines, which will print products as required in a model of edge-based manufacturing, with designs supplied and monitored through remote setups. Thus, new statements of design and manufacturing may be possible, with the fine print on goods reading *"genetically programmed in Uruguay, grown in India, to be consumed in Nepal and Bhutan."* Local-demand-supply-consumption models will become prominent in an already globalized world, with a marked emphasis on localized spatial circular economies. To ensure independence, assurance, and resilience, local manufacturing will be decentralized to several manufacturers, which will together compose a crowdsourcing production ecosystem. Smaller manufacturers will deliver items of their specialty to a larger assembling manufacturer, who will deliver the final products. It will be possible to handle orders quickly and dynamically, based on each stakeholder's timely capacity to deliver items, and robotized production plants will be able to operate remotely through virtualization. Managing the ecosystemic network of small manufacturers will utilize blockchain technology for supply chain management, smart contracting, and transactions. It will be possible to move production sites to new locations, enabling remotely controlled worksites and heavy-duty vehicles. Production models will be driven by sustainability, resilience, and the circular economy.

I5.0 will allow collaborative human–machine interaction (HMI) with robotization across services and industries, because human intelligence will be in perfect harmony with advanced cognitive computing. With real-time data, effective data monetization, and digital automation of the manufacturing process, businesses will be able to shift their focus to generating higher revenues from the servitization of products. Open interfaces and advanced manufacturing capabilities will help to overcome design complexities. With its ability to facilitate an extremely long tail of mass customization, it will also return control to customers in a haptic way. Furthermore, trustworthy quantum-enhanced I5.0 networking and services will provide the highest standards of safety and environmental protection.

The use of programmed organisms is becoming increasingly common in production. Genetic engineering and synthetic biology enable the creation of new kinds of organisms, as well as the modification of existing organisms for specific purposes in food production, chemical processes, textiles, and in the pharmaceutical and construction industries, for example. This will decouple growth from cost and resource usage. In interaction with hybrid bio-nano-artificial intelligence, industrial technology operators and maintainers will be forced to adopt a special synchronicity with these technologies, which in turn are adaptive to the workers. Unlike the Industrial Revolution of the 1900s in which the human became subjugated to the rhythm of the machine and prompted a backlash to the mechanistic life brought about by modern times, the new industry 5.0 technological revolution enabled by 6G technologies will bring about a new rhythm that links the biological dimension of the machine to that of the human. This will impact the human at sub-awareness levels, bringing about heightened nervousness, anxiety, and general discontent. However, unlike the early 1900s, humans will be unable to point to any immediate cause of the general malaise. The toll on psychological health may be enormous yet hidden in plain sight.

## Smart society (1C) Standardized experience & resource orchestration by society/ corporations

In the Smart society scenario, user experience is standardized, and resource orchestration is society/corporation-centric. Technology is developing rapidly, changing production methods and operating models. A growing number of things can be automated, production and operations can be decentralized, and interaction can take place remotely or via a virtual environment. This assumes continuous learning from individuals to keep track of development and evolve their professionalism. Making use of technology increasingly calls for changes in thinking models and operating methods. Gamification of working life may offer motivation for some people frustrated by the changes. Smart society builds dependable systems and communication in which standardized data is utilized by walled garden platform monopolies across verticals. The smart city focus is extended to rural inclusion. Multi-locality is the norm in combining city life and isolation from crowds. 6G will transform urban and rural living at the intersection of geopolitics, the growth of nationalism, rights to information transparency, and information citizenry.

A dependable communication system that allows remote work and telepresence in real-time mode will result in knowledge-based jobs and other net-based service sectors shifting to a bucolic life in which urban and rural life remains in healthy balance. The change in lifestyles will enable an emphasis on collaboration for the common good and making society more inclusive of the requirements of disparate cultures and sub-cultures. In this regard, there will be a marked shift to appropriate data and privacy regulation to support vested interests and motivations. In this smart economy, consumer insights, virtual finance, carbon-free consumption, low energy consumption, and global and fiscal sustainability will take centerstage. Thus, there will be an expansion of the social intangible economy, involving several types of online gaming, social media exchange, interaction in virtual holopresence interactions, and other forms of digital currency exchange. There will also be a rapid convergence of these various interactions, such as making online groups, communities, and institutional rules that will assist in creating an information citizenry and a reciprocal impact on real-world global issues.

The most important global concern will be to ensure mutual respect for people from every stratum of society. This will be possible through digital inclusion in all sectors, ranging from finance to education. The aim will be to create a just and egalitarian society through the use of appropriate information regulation and mutual distancing through the creation of safe and creative collaboration spaces that support the interests of like-minded groups. Actions at the level of individuals supported by 6G technologies will provide a morally sustainable world in which every citizen will be a self-aware informed citizen with a dual identity: recognizing allegiance to the nation, as well as living within the constraints of the global pan-dimensional virtually connected world.

## Communities (1D) customized experience & resource orchestration by society/corporations

In the Communities scenario, user experience is customized, and resource orchestration is society/corporations-centric. The sense of community created by 6G technology and the ability to directly collaborate with others enables humans to participate and act in society in an unprecedented way in countries where access to new technologies is the norm due to competences and skills in using new technologies. The sharing economy, crowdsourcing, and crowdfunding expand the space for new forms of organization and innovation.

In the heterogeneous society, social networks and the trust and reciprocity they foster will be highlighted from the perspectives of well-being and working life. Public network funding has traditionally been directed at unserved and underserved areas in terms of broadband access and coverage. Support for deployment programs will be extended to policy areas such as smart city community development, worksites, and ecosystems (such as harbors and airports), advanced health services, logistics and transport, public safety, and critical infrastructure at length. Hyperconnected and completely automated smart grids will be extended to a variety of vertical sectors, including electricity, the Internet, and healthcare, serving as a middle layer between humans and natural environments and enhancing the capabilities of both. These networks will have been assembled with a public-private-partnership funding model, with a view to sustainable growth and digital infrastructure usage. The human body will be a vital part of the Internet of Senses. Increased data will enable more personalized and preventive care in which AI-assisted analysis monitors personal indicators and compares them with larger population data, offering medical doctor consultation by a specific indicator trigger point. Biological processes and communication systems will be integrated with technical communication systems, providing online information about vital transactions and guiding us to take specific actions to remain healthy. In the event of infection, we will receive continuously updating diagnoses to be shared in real time with healthcare professionals, who will base their consultancy on AI-driven analysis. New treatments will also be developed based on genome editing and modifying the microbiome, for example.

Countries with less restrictive legislation will act as resource pools for corporations by providing cheap labor forces, natural resources, and (private) data about humans (use data, biodata, biological data, etc.). Frugal innovations will be developed to serve the growing customer base in low-income countries. Education powered by the Internet of Skills and the Internet of Senses will enable specialization from the school system's early grades. Learning will be tightly connected to personal data to react to any disturbance and ensure a successful study track based on individual interests. Students will be able to choose virtual courses and degrees from any university globally and visit digital twin campuses for interaction. Global networking during studies will support international career planning, which is done partly remotely.

A number of ethnic communities will have struggled to maintain their existence in everyday and virtual spaces. The nature of communities will change in varieties of ways and varying timescales: fragmentation of communities; dynamic tension between individuals and communities; the morphing of community values and identity; and other phenomena. Radicalized groups will have emerged, spreading terror both online and offline. The spread of cyberterrorism may affect every networked system in the world, resulting in a global crisis and a devastating effect on the world economy. In the wake of disasters (terrorist attacks, tsunamis, diseases, etc.), 6G technologies may also support the victims. The growth of human-body-powered networked devices will help a community to establish informational relations that aid the troubled, enabling the community to show resilient behavior and bounce back quickly.

At the level of communities, media interaction will result in the intensification of activities related to public opinion shaping. These will include the transmission of hate speech and fake news, which will also be experienced somatosensorily. This holistic experience of various forms of malevolence will have a much stronger impact than ever in mobilizing people toward crime and terrorism though virtual technologies. Special interest online communities will continue to proliferate. However, with 6G experiential technologies, these special online communities will move toward a more accelerated and hy-per-real set of interactions. Hate speech and associated activities will not only be symbolic but tangible. A final twist in the life of communities propelled by 6G technologies will be in terms of the "wisdom of crowds." In normal circumstances, this "wisdom" will allow for more egalitarian and informed decision making and empowerment. However, with the hyper-real experiential hate of 6G-enabled vitality, "wisdom" may be perverted without bounds, resulting in a bleak communal life.

### 3.2.2 Scenario theme #2 – Business

In the Business scenario theme, we chose the horizontal dimension to present value configuration and vertical axes value to capture logic, as depicted in Figure 6. The polar dimensions to the value configuration axes are traditional closed supply value chain focus and open ecosystemic-driven configuration. Value creation customer attraction and lock-in logic ends are the incumbent mobile operator-dominated model; the opposite is the expanded model with OTT, cloud, I5.0, and novel digital service provider stakeholders.

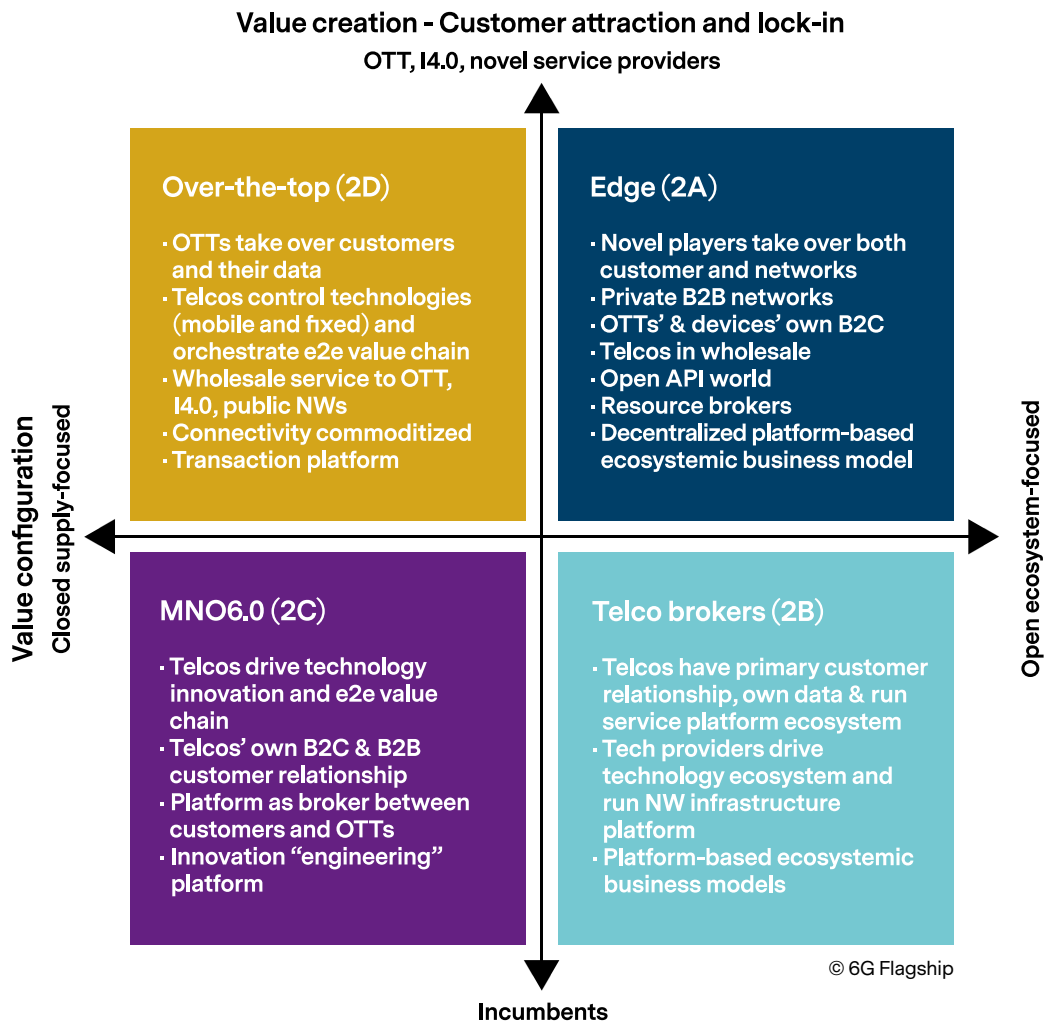

Figure 6. Scenarios based on value configuration and value creation

## Edge (2A) value creation by novel service providers & open ecosystem value configurations

In the Edge scenario, value creation is customer attraction- and lock-in-driven, and value configuration is open ecosystem-focused. This scenario stems from decentralized open value configuration and ecosystem-driven business models. Novel players have taken over both customer ownership and networks. OTTs and device vendors own the B2C customer relationship, while local tailored private cloud native networks attract the B2B customer. Telecommunications operators play a role as a wholesale connectivity service provider. Technology and innovation ownership are expanded, fully leveraging open API world and novel resource brokerage.

Furthermore, edge resources may be operated by local communities and special interest groups, e.g. in expanding services into remote rural areas, or universities and research organizations deploying their own edge resources to speed up local innovation. Nomadic sub-network edge elements will provide sustainable and ecologically efficient deployments. For example, communities that do not wish to have invasive technologies in their midst may hire edge-enhanced systems for occasional high-quality capture and streaming of local events. Banks, healthcare centers, and governance points may extend the regular telecom networks for the affordable inclusion of the masses. Onboarding, as well as billing of customers via digital cash and keys, can be done by these local entities, who in turn may pay the wholesale service providers for wider connectivity. These semi-autonomous 6G sub-network deployments will be heterogenous in nature, often encouraging innovative products and services. There will be specific network areas and zones. We can have a very personal zone through in-body communication applications producing data for daily diagnosis for individuals, or shared merely with a private medical doctor. The wider zone could be a common family network, shared strictly between family members at home and offering tailored services. Moreover, particular network zones will be shared with various interest groups enabling dynamic management based on personal preferences and the changing requirements of groups. Security, trust, and Identity management in such heterogenous edge deployments will be a challenge and opportunity for sustainable novel business models.

## Telco broker (2B) value creation by incumbents & open ecosystem value configurations

In the Telco broker scenario, value creation is driven by incumbents—the existing operators—and value configuration is open ecosystem-focused. Telco brokers have retained the primary customer relationship and have focused on monetizing data via the service platform ecosystem. Technology providers drive the technology ecosystem and offer an efficient network infrastructure via platform-based ecosystemic business models. The decoupling of technology platforms has lowered the entry barrier, allowing multiple entities to contribute to the innovations envisaged with 6G. Moreover, fine-grained modularity and open source allow highly specialized long tail solutions and services from smaller payers to be widely deployed, leading to innovation, and possibly to commoditization.

## MNO6.0 (2C) value creation by incumbents & closed ecosystem value configurations

In the MNO6.0 scenario, value creation is incumbent-driven, and value configuration is closed supply-focused. In this scenario, telecommunications firms drive technology innovation and control the traditional e2e value chain, owning the customer relationship in both B2C and B2B segments. It is strongly under MNOs' business-driven decisions that the advanced services enabled by 6G technology are available for various verticals. Key technology enablers utilized as the prerequisite for the decoupling of costs from growth are automated network slicing and leverage, and the use of higher frequency bands in conjunction with network densification. In addition to the technology innovation platform, they have established a transaction platform position between customers and OTT players. This tightly coupled deployment may provide optimal efficiency with respect to efficiency, environmental sustainability, and technology exploitation at the connectivity layer. Via the opening up of network interfaces, telecom firms co-develop within their value chain and open source software to address the long tail of specialized local and industrial use cases.

## Over-the-top (2D) value creation by novel service providers & closed ecosystem value configurations

In the Over-the-top scenario, value creation is customer attraction- and lock-in-driven, and value configuration is closed supply-focused. In this scenario, OTTs have taken over customers from telecom operators by utilizing their access to customer data. However, operators continue to control both the mobile and fixed connectivity technologies that are commoditized and orchestrate the related e2e value chain. Commoditized connectivity drives operators to create partnerships with OTTs, I5.0 service providers, and public networks and to provide wholesale services utilizing their transaction platforms. OTT players offer novel free or subsidized connectivity business models, utilizing revisited net-neutrality principles, affordably expanding their reach to the bottom four billion.

### 3.2.3 Scenario theme #3 – Sustainability

The sustainability crisis, which refers to the deterioration of the environment and exceeding Earth's carrying capacity, may significantly change our operating environment as we move to the 6G era. This scenario theme particularly recognizes the UN's SDGs as important drivers for 6G, but the approaches vary. Responding to increased environmental awareness requires changes in culture and practices, and has been accompanied by a polarization of views. Hybrid military, economic, technological, and cultural powers have become overlapping, exercising threats and hybrid influence. In the Sustainability scenario theme, the horizontal dimension represents power configuration. The vertical axis represents sustainability development (Figure 7). The polar dimensions opposite to the power configuration axis are centralized power and poly-nodal configurations. The ends of the sustainability dimension are the redefinition of the economy and its opposite, stagnation.

### Gaia (3A) Sustainability by redefinition of economy & poly-nodal power configurations

In the Gaia scenario, sustainability is driven by the redefinition of the economy, and power configuration is poly-nodal world-focused. Environmental awareness among people has increased and resulted in corresponding actions. Dissatisfaction with the current measures taken with respect to climate change and biodiversity has motivated a growing number of people to voice their opinions and participate in demonstrations. Instead of individual poles of power, the emphasis in global politics is on relationships and interaction. In addition to governments, other players, such as businesses, lobbyists, think tanks, international institutions and cities, activist organizations, play a significant role in this. In the 6G-enabled real-time economy, all the transactions between business entities are in digital format, generated automatically, and completed as they occur without store and forward processing. In innovating to zero, the cost of renewable energy and storage is falling. Energy production will also become increasingly decentralized as more and more people produce their own energy and sell what they do not need. In the circular economy, production and consumption are planned in a way that prevents waste from being generated, while materials and their value remain in circulation via sharing, leasing, repair, and reuse. In this context, deploying 6G could entail an application area to provide ultra-low-power communications through energy harvesting or wireless power to very small devices, for example. The counterforces to winner-takes-all monopolies include platform cooperatives, the peer-to-peer economy and sharing economic models, and the progress of the human-driven fair data economy and the fair distribution of wealth. This restorative economy will lead to a society characterized by broad empowerment, greater equality, a higher level of well-being, and better sustainability. The Internet of Senses and the Internet of Skills utilize advanced human–machine interfaces to enhance the human intellect and physiology toward transhumanism. In the Gaia scenario, societal resilience provides the ability to cope with and overcome adversities, the ability to learn from past experiences and adjust to future challenges, and the ability to craft sets of institutions that foster individual welfare and sustainable societal robustness to future crises.

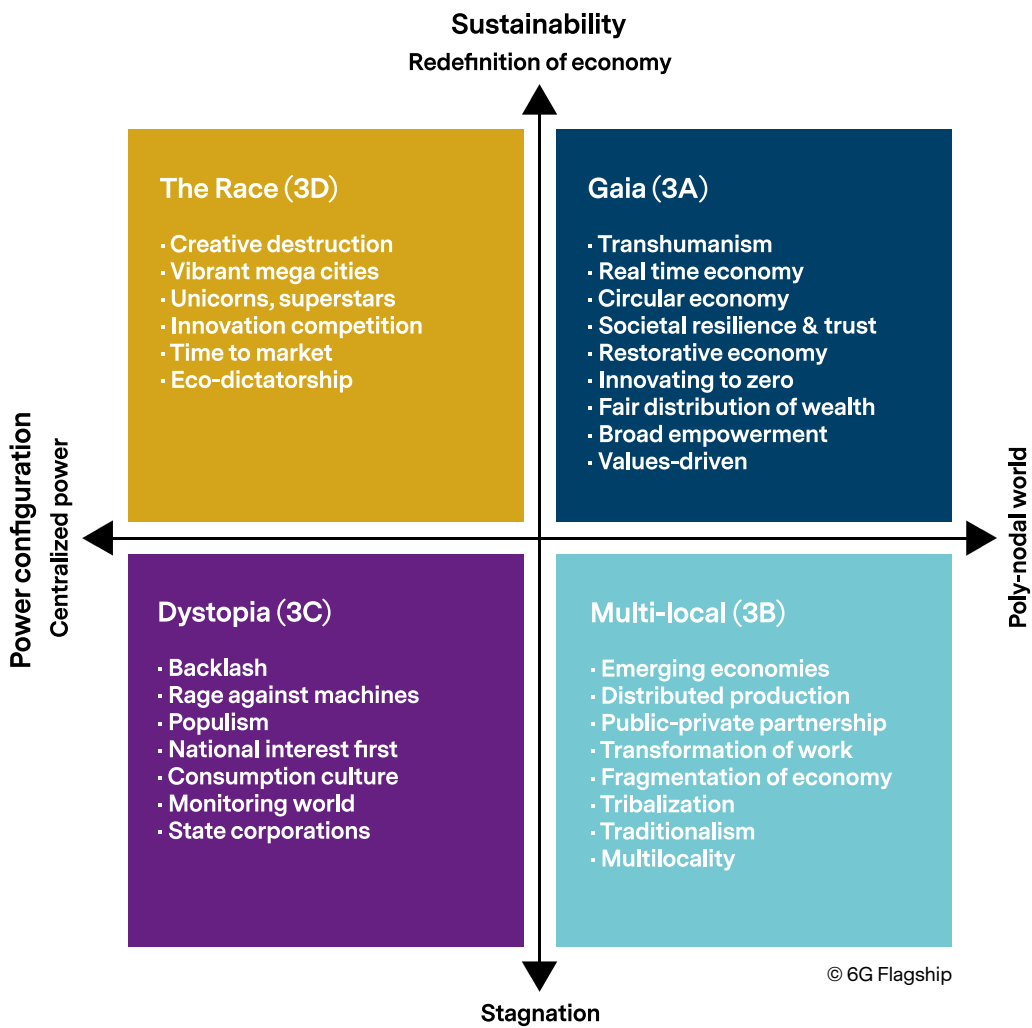

**Figure 7. Scenarios based on sustainability and power configuration**

## Multi-local (3B) Sustainability by stagnation & poly-nodal power configurations

In the Multi-local scenario, sustainability is stagnation-driven, and power configuration is poly-nodal world-focused. A shift has happened in global politics from a multipolar to a poly-nodal world. The geopolitical power blocs—the US, Europe, and China—give way to a networked world with nodes comprising countries, emerging economies, corporations, and other non-state actors. In the face of disruption, people turn to increasingly polarized tribes and bubbles formed around values, place of residence, political opinions, consumption choices, or lifestyles for guidance. The fragmentation of the economy, transformation of work, and new organizational models of the sharing and platform economy have challenged the traditional relationship between the employer and employee, and what the benefits are. Working life is becoming increasingly diverse, and there is an emphasis on ensuring people's livelihood and competence building. Universities offer tailored virtually augmented education environments to enterprises and private entrepreneurs. Consumers favor domestic or local products, and can themselves choose where and how goods and products are manufactured. 3D printing allows many products to be manufactured at home or in the neighborhood. Distributed local production is on one hand practiced by enterprises using a network of geographically dispersed manufacturing facilities coordinated using 6G; on the other hand, it is practiced by local manufacturing by prosumers. Traditionalism is developed as a response to disorder and favors public–private partnership.

## Dystopia (3C) Sustainability by stagnation & centralized power configurations

In the Dystopia scenario, sustainability is stagnation-driven, and power configuration is centralized power-focused. We continue to inhabit a consumption culture in which nature is seen as a free resource that we use as we wish. The wealth generated by economic growth is not distributed sustainably and is concentrated in the hands of a shrinking minority. Occasional large disasters do not make wealthy people act if they are not threatened themselves. Weakened future prospects, the fragmentation of the political map, and the polarizing effect of social media have led to a rise in populism, which emphasizes the division between the elites and the masses. The benefits of internationalism are not acknowledged, because they are considered too indirect, and its negative aspects are emphasized in the discussion. Globalization has also led to an opposing reaction in the form of increased nationalism and an emphasis on national borders, and favors state corporations. Democracy is challenged by ideas of practical autocracy and technocracy, as well as by the notion that democracy is too slow or ineffective to respond to the urgent questions of our time. The need for rapid major changes and a yearning for simple solutions has made strong leaders more popular, presenting a challenge to individual freedom and democracy. The amount of disinformation is growing, and efforts to influence opinions are increasingly geared toward instigating confusion and discord. A digitalization backlash has happened, and people are raging against machines.

## The Race (3D) Sustainability by redefinition of the economy & centralized power configurations

In the Race scenario, sustainability is redefined economy-driven, and power configuration is centralized power-focused. The urgency of climate and sustainability action has led to eco-dictatorship and creative destruction, because a voluntary change in people's behavior is considered so unlikely. The population becomes concentrated in a small number of growth centers, where vibrant megacities and unicorn superstars dominate innovation addressing individual technologies and the ecosystems they form. The process of innovation competition incessantly revolutionizes the economic structure from within, giving time for market advantage.

### 3.2.4 Scenario summary

Twelve alternative future scenarios were designed under three embedded scenario logic themes: User experience, Business, and Sustainability, depicted in Figure 8. User experience can be seen as a sub-set of Business in the middle of Figure 8, and Business is a sub-set of Sustainability, forming the widest contextual level for all scenarios. To summarize the discussion of 6G business scenarios, the team proffered both optimistic and pessimistic scenarios, with various reasons for optimism and pessimism by scenario. The scenarios were assessed at the end of the workshop series. First, the likelihood of the created scenarios coming into being was assessed. The *probability* of the scenarios arising was evaluated against the identified forces influencing them. Next, the *plausibility* of the scenarios was assessed, based on their coherence, by examining the potential alternative futures for 6G business events that could occur within their assessment. The third assessment step was to identify which scenarios were the most preferred within teams. The *preferability* assessments of the scenarios were based on the values and choices the teams made regarding alternative futures. The teams appraised the different scenarios in a largely similar way. Probability and plausibility were assessed as clearly correlating; the preferable scenarios were only considered probable or plausible in a few cases, although they were seen as something which should be sought.

Both the most probable and most plausible scenarios stem from evolutionary supply-driven trends toward a multi-local networked world based on strong trends with low anticipated uncertainty. The most probable and plausible business scenarios, OTT (2D) and MNO6.0 (2C), build on the balance between competition and protective market views. In the Sustainability themed scenarios, the Multi-local scenario sharing both the dystopic and utopian themes is seen as simultaneously the most probable and plausible. All the preferable scenarios, Gaia (3A), Edge (2A), and Customer6.0 (1A), represent revolutionary demand-driven transformations toward sustainability, empowerment, and open ecosystems. They are based on high impact forces with higher uncertainty compared to most probable and plausible scenarios.

In the preferred 6G future, automatic collection of different kinds of data from humans, our environment, and its analysis are used for highly sophisticated products and systems that make people's lives easier and provide better user experience through convenience, because everything is automated. The edge resources will be operated by local communities expanding services to remote rural areas, or research organizations deploying their own edge resources to accelerate local innovation. 6G will enhance platform cooperatives, the peer-to-peer economy and sharing economic models, and the progress of a human-driven fair data economy and the fair distribution of wealth. To summarize, we identified drivers, barriers, and challenges regarding the choices for developing the preferred 6G business future, depicted in Figure 9. These drivers, barriers, and challenges could concern all stakeholders in future 6G business. Key transformative global drivers concern climate change, sustainable development goals, and decentralization toward a networked poly-nodal world. External barriers to preferable future scenarios are uncertainties related to power of dominating platforms, AI and HMI rights, and the regulation of resources. Key internal challenges were identified as building a disruptive business model leveraging sharing economy antecedents while coping with the empowered users' rights.

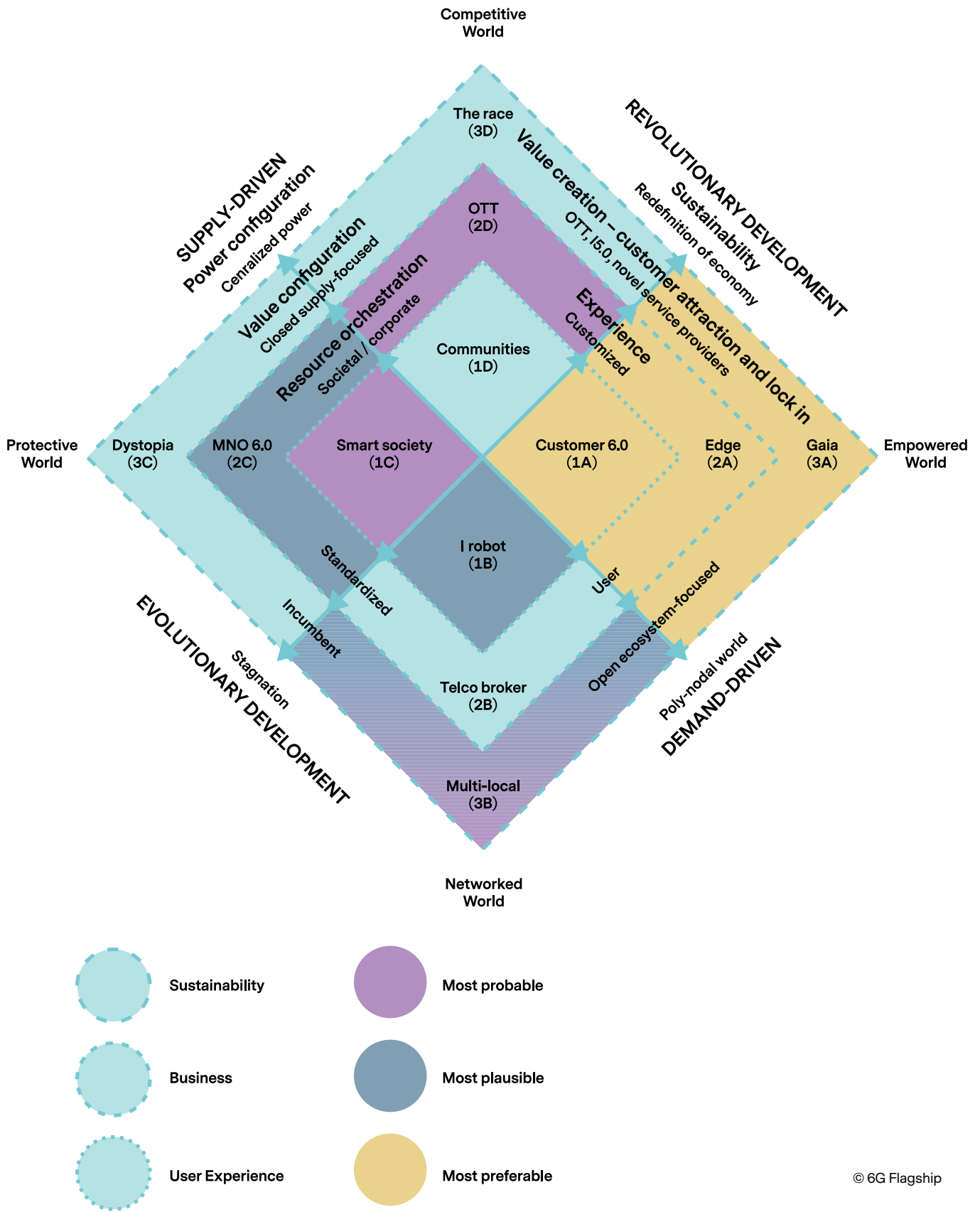

Figure 8. Summary of three scenario logic themes: User experience, Business, and Sustainability

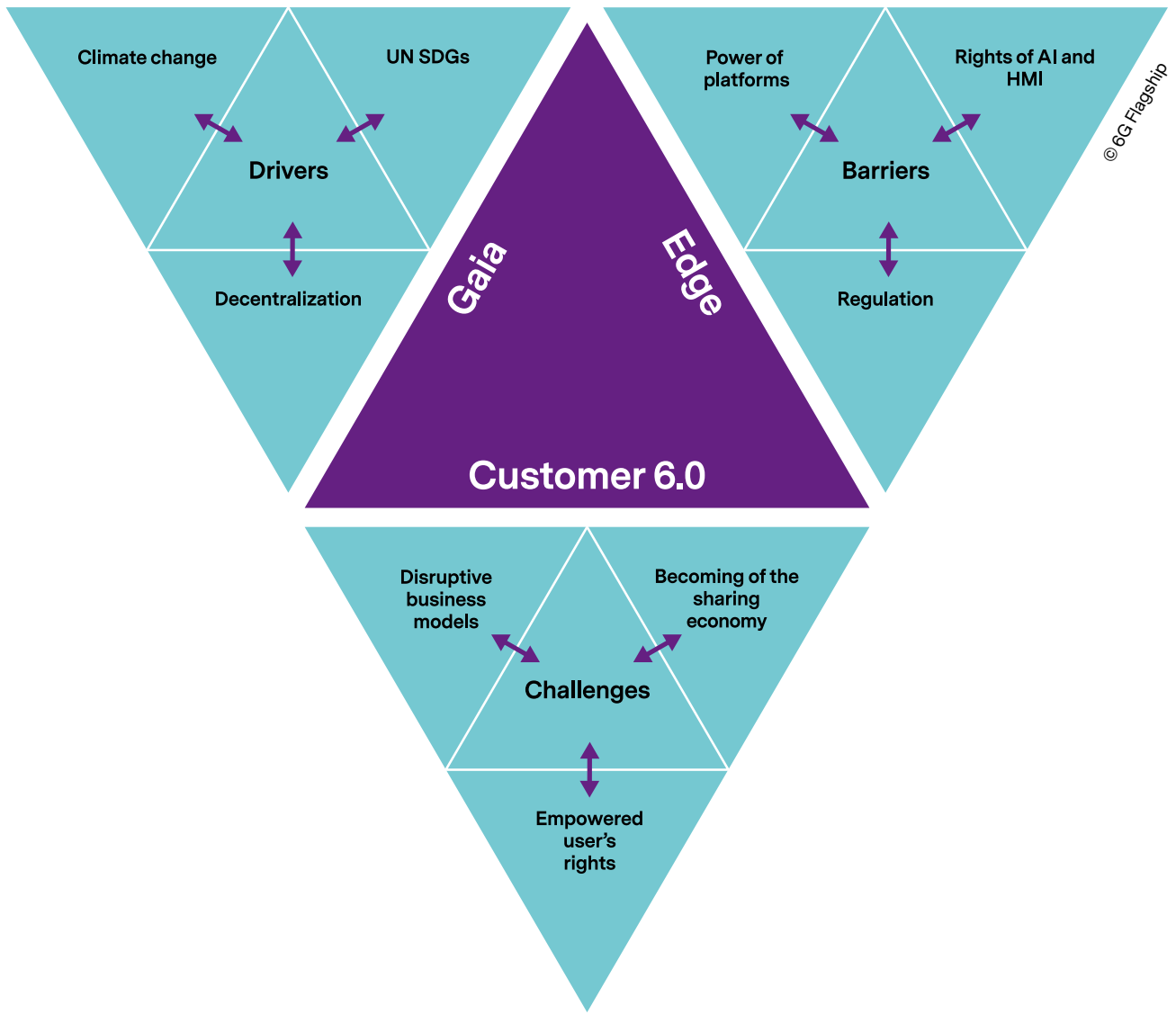

**Figure 9. Essential choices for developing preferred 6G business futures**

# Implications and options

**4**

## 4.1 Scenario backcasting to key technology trends and enablers

The created scenarios were backcasted to key strategic technology trends identified in the scenario process and outlined in technology-themed 6G white paper workshop results (6G white papers, 2020). In backcasting, the team worked backward to identify the key 6G technology trends and their technology enablers required to reach the created preferred future scenarios from the current state. Identified 6G technology trends and enablers, and positioned utilizing network architecture domains (Ziegler & Yrjölä 2020) and the technology development phase are summarized in Figure 10.

Key technology enabling themes to be explored will include the pervasive leverage of machine learning and artificial intelligence across architectural domains – namely, to flexibly define air interface and optimize service management and orchestration in 6G "network of networks" topology and the platform ecosystem. THz research, in the sense of utilizing spectral bands of above 100 GHz for both communications and multimodal sensing purposes, is one of the prominent topics, enabling connectivity data speeds in the Terabit/s range. We foresee millions of sub-networks (and devices becoming the network) in conjunction with extreme performance attributes in terms of sub-millisecond latency, high reliability, time-sensitive determinism, and advanced ways to assure security, privacy, and trust. In a future flexible cognitive network with self-optimizing radios, AI and ML can be used in concert with radio sensing and positioning to learn about the static and dynamic components of the radio environment, predict link loss events at high frequencies, proactively decide on optimal handover instances in dense city networks, and identify the optimal network spectrum and radio resource allocation for base stations and users. With a strong sustainable focus on human needs and global inclusion, non-terres-trial networks (NTN) extend to uncovered areas of Earth and space. Furthermore, GEOs, MEOs, LEOs, HAPS, and drones will be used for extreme 6G system reliability and resilience for public safety and critical infrastructures. Sub-networks, edge cloud computing and extended devices consisting of multiple local entities in unison will enable a local and instant information service, e.g. for a fast discovery of people, services, devices, resources, and any local information near the user that cannot be collected by centralized search engines. Such an edge intelligence and information service platform could be used, e.g. in the creation of a highly local and dynamic marketplace for content, services, resources, and information. Edge computing capabilities address specific service demands, including bandwidth management, latency, sensitivity, security and privacy, local control and service continuity, analytics and digital automation, and support for constrained environments and energy efficiency. Digital trust, enabled by quantum computing and distributed ledger technologies like blockchain and smart contracts, provides business in a secure and predictable digital society with world-class cyber security, public safety, and fintech solutions. Cloud and network providers' cloud native edge computing may become natural central points, representing the source and destination of much of the demand, combined with context analytic-enabled optimization capabilities. This may create a new competitive advantage over centralized OTT services through content caching, optimized local content distribution, location services, video analytics, and AR and IoT application for existing customer segments, as well as for new vertical business segments. Furthermore, autonomic security functions will be increasingly applied at the network edge to protect the network by reacting locally to threats. Hierarchical security and privacy methods will allow networks to monitor and adjust network slices, virtualized elements, and sub-networks in accordance with detected security threats without human intervention.

© 6G Flagship

| Network architecture domains | | | |
|---|---|---|---|
| **Security & privacy** | Local edge and sub-network security | Digital trust | Security autonomics |
| **Digital value platforms** | Human–machine interface | Augmented Intelligence Digital twins | Holopresense |
| **Cognitive network management** | Resource brokerage & open APIs | Zero human touch cognitive MANO | Context-aware network orchestration |
| **Core network** | Cloud native | Hyper-specialized virtualization & slicing | Ubiquitous sub-nets |
| **Local edge cloud** | Caching and edge computing | Time-sensitive critical local intelligence | Ubiquitous augmented cognition |
| **Access** | Private local virtual open access | Non-terrestrial networks | Self-optimizing comms and multi-sensing |

Technology trend and enabler evolution

**Figure 10. Backcasted key 6G technology trends and enablers by architecture domain**

In the cognitive network management and orchestration (MANO) layer, there are intelligence needs in the self-configuration, optimization, and orchestration of virtual resources and sub-networks to meet highly specialized dynamic content, and contextual and event-defined needs. A programmable network will utilize a digital twin as an exact digital replica of complex physical assets, processes, and systems, providing a detailed understanding of how the real system is behaving and predicting what it will do next. The resources and assets required to meet the network's versatile needs will then be provided by different stakeholder roles providing physical infrastructure (facilities, structures like street furniture, internal walls), equipment (devices, networks), data (content, context) under the regulatory framework set by policymakers. Demands and resources are brought together through matching/sharing stakeholder roles, including different kinds of operator (local or vertical-specific, fixed, mobile network/satellite operators), resource brokers, and various service/application providers such as trust/security providers. This is supported by an open accelerated platform extended from elements and interfaces toward data and algorithms leveraging IT practices and the reuse of their large open source heritage.

6G technology will provide digital value platforms for the delivery of advanced extended reality applications, supporting immersive mobile media experiences that extend over the entire continuum of digital computer-generated virtual worlds. A key emphasis in the growth of digital value platforms will be the convergence of multimodal engagement with media and the physicality of lived experience (embodied interaction, mixed-media interaction, tangible interaction, among many other interaction design-based computing approaches). The related human-centered visual, audio, and interactive computing applications are seamlessly accessible through human–machine interfac-es engaging the five senses of sight, sound, touch, smell, and taste. Individual and collaborative holographic presence with virtual 3D sound environments will revolutionize multimodal media experiences and communication practices, including real-time virtual 3D teleportation. Users will be able to seamlessly switch between any form of immersive mobile extended reality encompassing virtual reality (a completely virtual world), augmented reality (an enriched real world), and mixed reality (virtual and augmented objects). The sense of touch in immersive mobile digital realities will extend well beyond conventional external devices like gloves, shoes, and joysticks but will allow for much more sophisticated haptics that essentially engage the entire body. In view of the great complexity associated with a human-centered ecosystem with respect to interconnected immersive digital reality applications, a variety of visual and interactive computing technologies will be employed, e.g. computer graphics, virtual acoustic and reverberation effects, advanced haptics, mobile digital reality-enabled devices, and 6G connectivity. It will be possible to identify applications of these digital realities in every vertical industry, including advertising, architecture, the automotive industry, design, education, engineering, healthcare, interior design, libraries, marketing, media, medicine, music, news, real estate, retail, sports training, television and film, and travel. These immersive digital realities will facilitate novel ways of learning, understanding, and memorizing subjects in many sciences like chemistry, physics, biology, medicine, and astronomy. It will be able to move customers back in history, dive into the actual artifacts of ancient times, and experience related stories or propel them into the future to explore exciting new worlds. Healthcare will benefit from mobile digital realties using holoportation, along with novel haptics for both diagnosis and treatment. Real-time shopping experiences will allow immersion and presence in a virtual store, including interaction with products, to eventually perform an actual purchase.

## 4.2 Strategic options

The range of opportunities in the disruptive solutions described in the preceding chapter raises the question of how to commercialize them, and who is in a position to do so: incumbent operators already present in the mobile communications ecosystem; new entrants from other ecosystems; or startups developing novel business models. To answer these questions, 6G business drivers, barriers, and challenges and the opportunities of the selected business scenarios were further analyzed utilizing the simple rules strategy framework (Eisenhardt & Sull 2001). Simple rules strategic options, including opportunity, how-to, boundary, priority, timing, and exit-related rules, were created for the most plausible MNO6.0 scenario (2C) and the most preferred Edge scenario (2A).

### Simple rules for the most plausible MNO6.0 scenario

Simple rules for the traditional mobile network operator evolution in the MNO6.0 scenario (2C) are depicted in Figure 11. The opportunity for businesses can simply be seen to lie in the utilization of MNOs' wide existing customer base, with a growing demand for capacity. Fast access to a new wideband spectrum, preferably avoiding coverage obligations, will enable the continuation of the utilization of a dominant position in the markets. Dynamic flexibility, shortened time-to-market, and massive cost optimization will be enabled by leveraging a paradigm of automation and dynamic instantiation of thousands of slices on demand. An automated marketplace of resources, services, and performance attributes will be enabled. ML/AI-based analytics and cognition will be driven by metadata architecture and federation, and pre-dictive triggering of corrective measures from intent/usage-based prediction of new services/business needs will become the norm. In such a scenario, the network's verifiability and trustworthiness will require security guarantees and proofs. Service-driven network management and automating-related network management tasks for efficient and flexible extraction of value are likely to be at the heart of MNO6.0. Dynamic service provisioning will be a key enabler in automating the monetization of available performance capacity. For spot markets of resources with certain performance attributes, automated service management is expected to be the key enabler across the network, with full KPI granularity. In some locations, it will be possible to use enhanced small cell deployments to provide premium quality of service (QoS) connectivity. All investments should be made to strengthen customer lock-in and dominant market position in connectivity, and as regulation allows, enhanced with customer data. It will be possible to achieve this by acquiring all the available spectrum if possible, and combining existing and new spectrum assets to deliver high data rates. Becoming a wholesale and hosting platform provider for other operators could also enhance the utilization of the dominant market position. To strengthen market boundaries, dominators could leverage their installed base, existing sites, and Third Generation Partnership Project (3GPP) evolution in the build-up of spectrum-sharing-based asset businesses. MNO6.0 could also attempt to convert novel connectivity service providers into mobile virtual network operators (MVNOs). Direct contact with the regulator is required to remind the regulator that earlier investments in the networks have sufficiently long payback periods. A supply-side platform-based regulation logic is called for and is lobbied as a way to define and provide the required level of privacy and security for users.

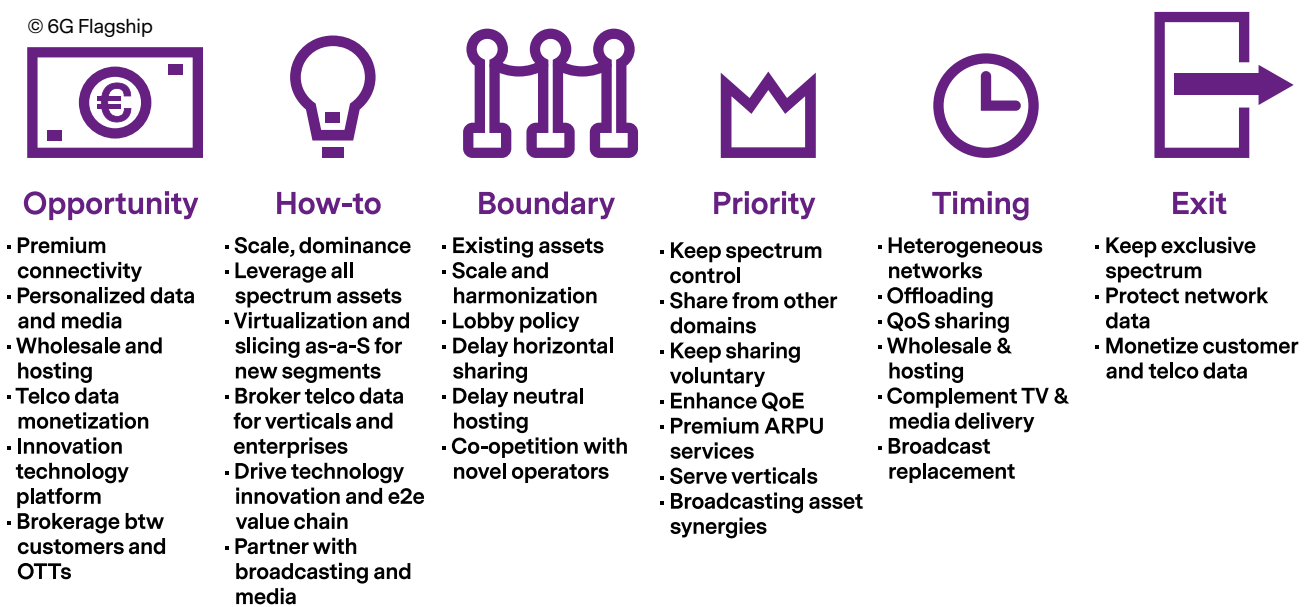

© 6G Flagship

| Opportunity | How-to | Boundary | Priority | Timing | Exit |
|---|---|---|---|---|---|
| · Premium connectivity<br>· Personalized data and media<br>· Wholesale and hosting<br>· Telco data monetization<br>· Innovation technology platform<br>· Brokerage btw customers and OTTs | · Scale, dominance<br>· Leverage all spectrum assets<br>· Virtualization and slicing as-a-S for new segments<br>· Broker telco data for verticals and enterprises<br>· Drive technology innovation and e2e value chain<br>· Partner with broadcasting and media | · Existing assets<br>· Scale and harmonization<br>· Delay horizontal sharing<br>· Delay neutral hosting<br>· Co-opetition with novel operators<br>· Lobby policy | · Keep spectrum control<br>· Share from other domains<br>· Keep sharing voluntary<br>· Enhance QoE<br>· Premium ARPU services<br>· Serve verticals<br>· Broadcasting asset synergies | · Heterogeneous networks<br>· Offloading<br>· QoS sharing<br>· Wholesale & hosting<br>· Complement TV & media delivery<br>· Broadcast replacement | · Keep exclusive spectrum<br>· Protect network data<br>· Monetize customer and telco data |

**Figure 11. Simple rules for the traditional mobile network operator in the MNO6.0 (2C) scenario**

Regarding decision priorities, retaining control of network technology, the related spectrum, and cognitivity in the network will be essential. To serve verticals, edge computing will become a new control point. Local sharing with other web-scales and OTTs could follow the potential for internal asset leverage that has been reached to attain media content and industrial data. Regardless of the utilized business model, MNO6.0 should never give up the spectrum and customer data—even if not fully utilized.

## Simple rules for the preferred Edge scenario

For the novel 6G operator in the Edge scenario (2A), the opportunities are quite different from those for MNO6.0, as is depicted in Figure 12. The opportunity lies in the utilization of new, local, and specialized demand, challenging incumbent MNOs in narrow business segments specializing in governmental, municipal, vertical, or enterprise customers, or differentiation with special mobile devices, mission critical communication, or web-scale cloud-based services. Furthermore, vertical differentiation may create an advantage in specific industry segments like education, healthcare, manufacturing, logistics, mining, energy, media and entertainment, and eSports. 6G users will expect future interactions to be built around human-centered end-to-end experience as the dominant concept. In addition to the human–machine–interface with networks and devices, empowered "experientials" will value access to "Internet of Skills" expertise, and the ability to evaluate and reflect on the technologies and their role in the user's own life, as well as more widely in society. Experientials' experience of being able to influence things and sense that they are important will be essential in all forms and work sectors, whether they revolve around entertainment or factory work. For a company to succeed, the questions asked will be in terms of "what makes our process unique for experientials?" Managers will be involved in addressing the boundary rules required to focus on which aspect of the user experience they will cater to or assign ranks to various modes of the customer experience. They will also require a set of timing rules to tune into the market and exit, based on their anticipation of key customer experience modalities.

Thinking and acting locally, close to the customer, promoting resource sharing, and utilizing the lowest cost spectrum and virtualized cloud infrastructure may open opportunities to scale up from local operations to a multi-locality business. The benefits of deploying private networks will include: security and data control, with full separation from public networks; access to premium services in locations not reached by public networks; extreme flexibility, scalability and customization via novel orchestration interface and automation; trustworthy reliabilities and latencies, and autonomous deployment planning of cluster-based and dense networks with node-level intelligence; and predictive analytics of system reliability and link quality. ML/AI will be a pervasive enabler of such scenarios of cognition. Furthermore, predictive analytics could be used to adjust physical layer communication and wireless networking cooperative protocols. Access points and distributed units will act as a huge distributed sensor to locate users and machines with very high precision and obtain clarity about their physical states. The option of fusing other sensors such as optical and sound sensors in such a scenario will greatly contribute to augmented reality and digital twin visions. Organizations will be able to build and operate their own sub-networks, and buy solutions from equipment vendors, systems integrators, cloud providers, or as-a-service from a wholesale provider. Networks are likely to be extreme-edge and edge-centric, as well as data flow-based across the network. Workloads will be dynamically scheduled for different tiers in the hierarchy of data centers across the network topology. Network functions and service function chains will be assigned dynamically, based on the optimal balance between consumed and available resources, connectivity and latency requirements, and energy consumption targets through online multi-object optimization algorithms. Service discovery must operate at the transaction time level to match the changing context and resource allocation situation across distributed cloud facilities: this may lead to the refactoring and distribution of network functions. It will be possible to deploy networks as standalone (sub-)networks or integrated/semi-integrated, with existing public operator networks playing a complementary role in the markets. This will call for a multi-sided platform-based regulation that goes beyond spectrum regulation to govern the more important privacy and security of users, even in rules based on community governance. Some will build the network but outsource operation, maintenance, or support. They could also adopt a boundary-reinforcing role by seeing local regulation as an opportunity and seeking to convert service fragmentation into a source of competitive advantage. Regarding decision-making priorities, edge must pay attention to customer value and quality of experience (QoE). By focusing on operational efficiency, zero-touch slicing and automation, and investment minimization, they could attempt to do business where and when a low-cost spectrum commons, local licensing, or sharing is freely available. As an exit plan, edge could consider becoming a virtual operator, focusing on customer ownership and the control point at the edge.

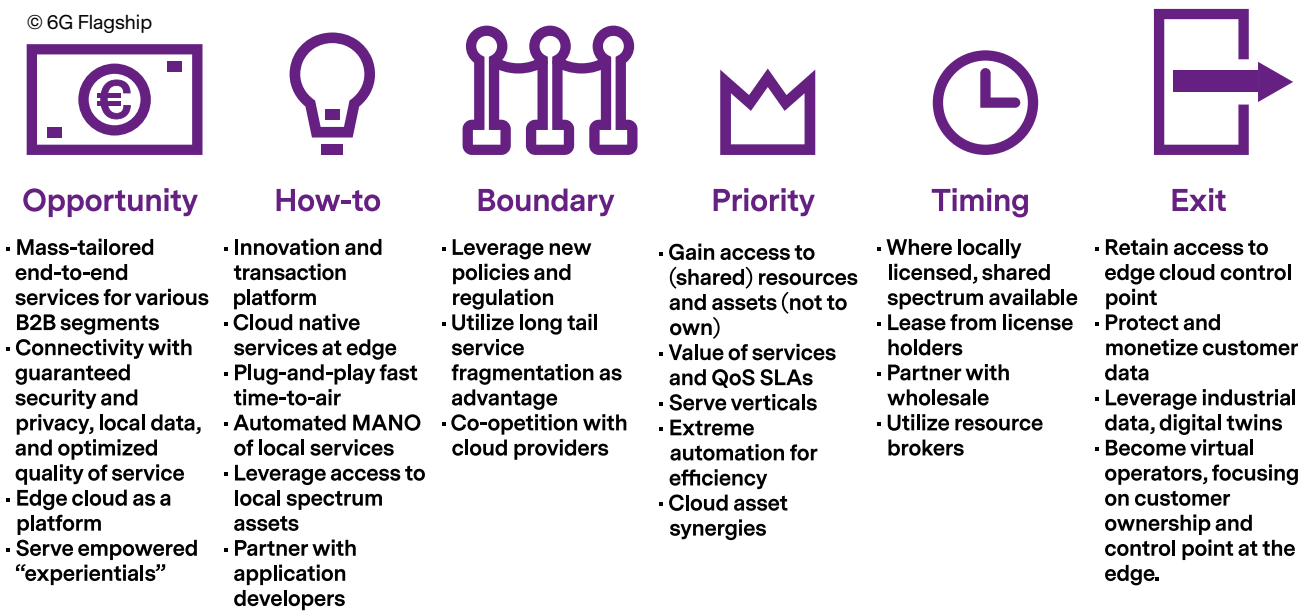

© 6G Flagship

| Opportunity | How-to | Boundary | Priority | Timing | Exit |
|---|---|---|---|---|---|
| • Mass-tailored end-to-end services for various B2B segments<br>• Connectivity with guaranteed security and privacy, local data, and optimized quality of service<br>• Edge cloud as a platform<br>• Serve empowered "experientials" | • Innovation and transaction platform<br>• Cloud native services at edge<br>• Plug-and-play fast time-to-air<br>• Automated MANO of local services<br>• Leverage access to local spectrum assets<br>• Partner with application developers | • Leverage new policies and regulation<br>• Utilize long tail service fragmentation as advantage<br>• Co-opetition with cloud providers | • Gain access to (shared) resources and assets (not to own)<br>• Value of services and QoS SLAs<br>• Serve verticals<br>• Extreme automation for efficiency<br>• Cloud asset synergies | • Where locally licensed, shared spectrum available<br>• Lease from license holders<br>• Partner with wholesale<br>• Utilize resource brokers | • Retain access to edge cloud control point<br>• Protect and monetize customer data<br>• Leverage industrial data, digital twins<br>• Become virtual operators, focusing on customer ownership and control point at the edge. |

Figure 12. Simple rules for the novel 6G operator in the Edge (2A) scenario

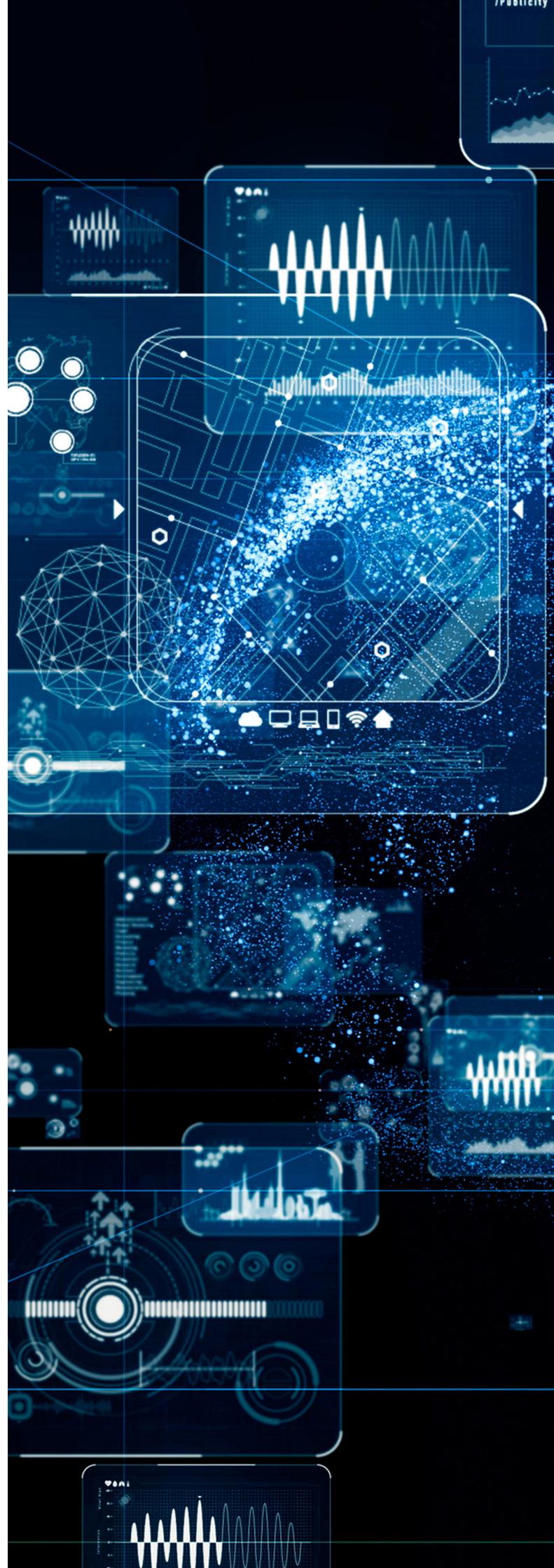

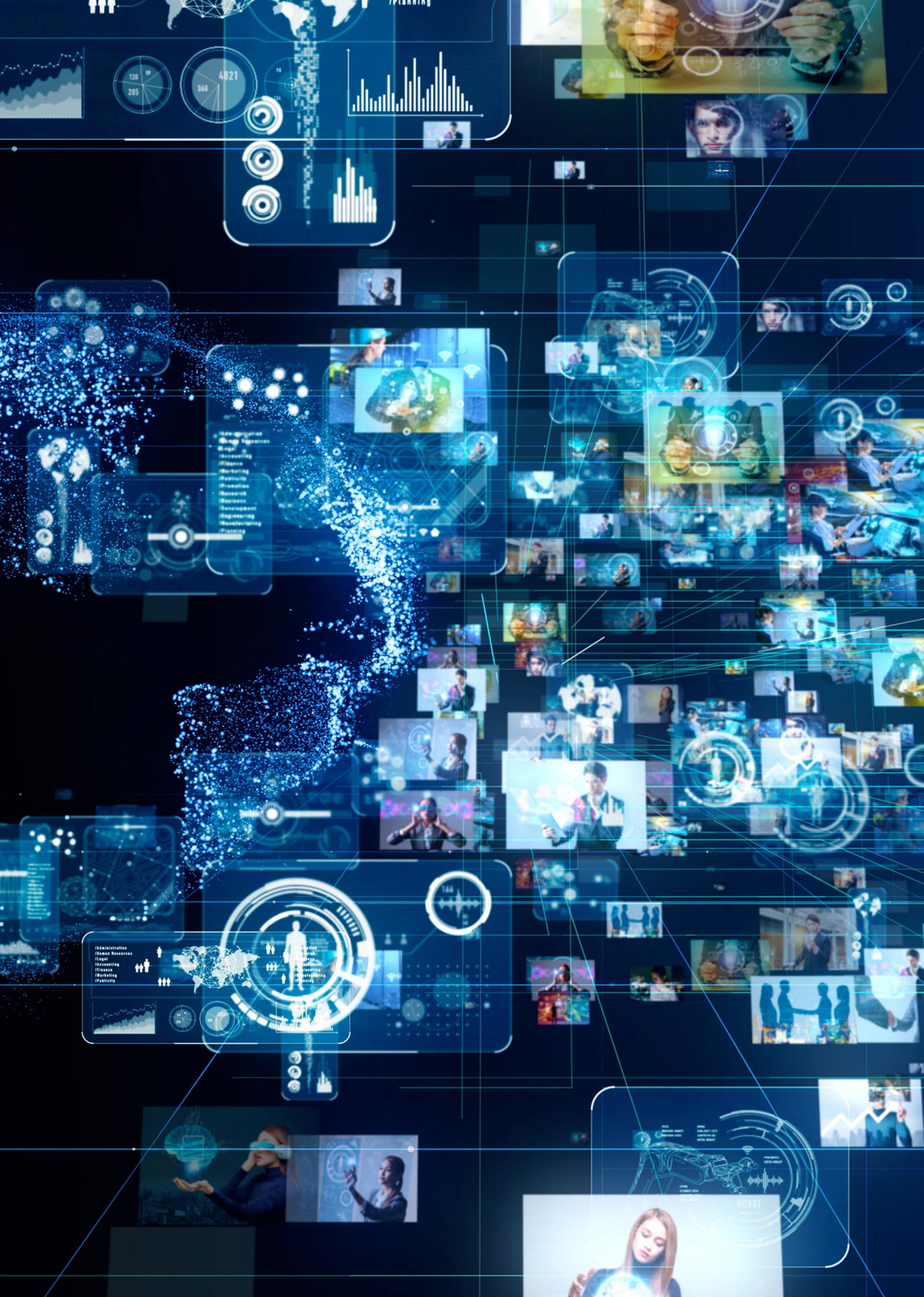

# Discussion and future research topics

# 5

Next, the developed scenarios are analyzed and positioned, utilizing the integration of the ecosystem-focused business model configuration and open value configuration framework (Xu 2019), as summarized in Figure 13. Gaia, Edge, and Customer6.0 scenarios can be seen as transforming business toward an open ecosystem, focusing on configuration stemming from the customized user experience, attraction and lock-in by OTTs, I5.0, novel service providers and building on sustainability, empowerment, and redefined economics values and networked poly-nodal decentralized power configuration. The centralized and community-based OTT, Communities, and Race scenarios utilize mixed value configuration and demand focused business models while retaining the value capture focal point. The protective MNO6.0, Smart society, and Dystopia scenarios build on a closed supply-focused value and power configuration. These incumbent-power-based stagnated scenarios evolve toward open demand-focused networked business in the Telco brokers, Multi-local, and I Robot scenarios.

In analyzing the scenarios, we find the main forces, trends, and uncertainties stemming from the five common meta trends identified by Sitra (Dufva 2020): *the need for ecological reconstruction; the strengthening of relational power; the aging and diversification of the population; the redefinition of the economy; and technology is embedded in everything.* Furthermore, the scenarios at each level—User, Business, and Sustainability—can be seen as parallel, because they do not exclude the existence of the other scenarios. An additional insight can therefore be gained by cross-examining their content, linkages, and tensions at different levels such as markets, verticals/industries, ecosystems/networks, business models/strategies, products/services, customers/segments, and different types of user. This cross-examination reveals a consistency in the importance of privacy and security issues related to business and regulation needs across all scenarios: public/governmental; corporate; community and user(s) perspectives and aims of governance; ecosystem configuration related to users, decentralized and community business models and platforms; user empowerment; and the role of 6G service location specificity. To summarize, we identify five key perspectives and themes as indicators and signposts that can provide an advanced insight into how or which of these scenarios will actually unfold. The first concerns *time-to-market* in 6G through monitoring which of the businesses in Figure 13 will emerge first. The second concerns *complementarities*, potentially large spill-over effects, and the competitive landscape within the industry, especially to observe the platform wars expected to emerge. The third concerns the analysis of the *appropriability* regime, bottleneck assets, and who profits from innovation and how. The fourth assesses the antecedents of the *sharing economy.* The fifth concerns the recognition and deployment of *sustainability goals* and the redefined economy.

The research's practical implications are twofold. First, the twelve alternative integral business scenarios presented can be used as a baseline for conceptualizing potential novel ecosystemic business models for the emerging 6G business ecosystem. Second, the simple rules strategic framework utilized to deepen scenarios will allow ecosystem stakeholders to create strategic options and indicators in exploring novel 6G opportunities. The previous chapters indicate a multitude of alternative future business opportunities and models for different 6G ecosystem stakeholders. We therefore propose a focus on business models as a way of thinking and ecosystem stakeholders' business model-related choices regarding opportunities, value-add and capabilities, and their expected consequences as scalability, replicability, and sustainability. With the right business choices, opportunities will be identified related to novel and unmet needs, new types of customer and service provider, as well as the interfacing of humans with machines.

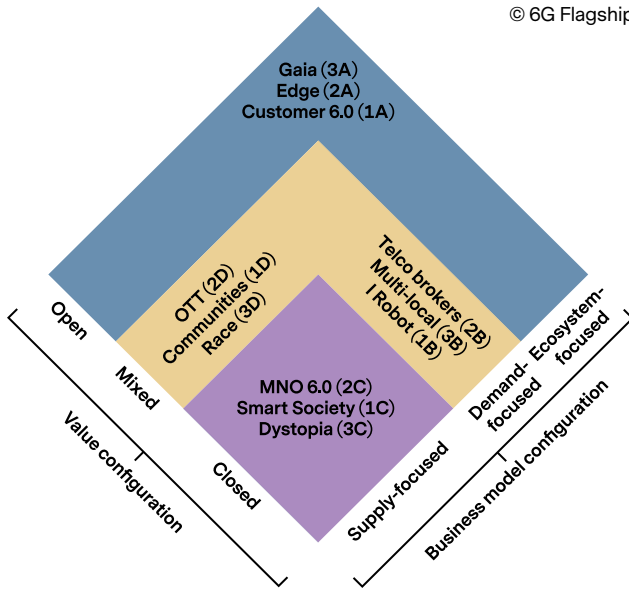

**Figure 13. Scenario positioning in the integration of ecosystem-focused business model and open value configuration. Framework adapted from Xu (2019).**

New value-added is seen to come from real-time and trustworthy communications, local data and intelligence, and the commoditization of 6G resources as its competitive advantages, including extreme capacity and security, transaction and innovation platformization, and ubiquitous access. The expected business consequences of scalability may be related to the long tail of services, dataflow architecture, automation, and open collaboration between partners; in terms of replicability, to deliberately design modularity and complementarity within platforms; and in terms of sustainability, to empower users and communities, and the utilization of sharing economic mechanisms in the markets. Figure 14 summarizes these key elements of business model thinking.

The scenario, simple rules, and business model analysis direct attention to the dynamics of 6G platforms and ecosystems. Business models as enablers of ecosystemic interaction and open value configuration is well established in the business model literature (Gomes et al. 2018; Xu 2019). However, digital convergence across industries and multi-level 6G platforms and ecosystems is creating a complex strategic environment that can lead to incomparable and distinct opportunities, as well as emergent problems. In particular, unanswered questions remain about ecosystemic business models in the context of sustainability. According to our findings, business ecosystems that aim to bring together stakeholders to solve systemic sustainability problems will require open ecosystem-focused value configuration and decentralized poly-nodal power configuration, focusing on the long tail of specialized user requirements that crosses a variety of industries. Furthermore, based on the results of the simple rules strategies, opportunity-driven business models of the novel digital service providers serving the customer of the future are seen as a sound point of departure for discussion about designing the ecosystemic business model.

To achieve an actionable 6G future, types of multi-level platform interaction need to be reconsidered, because it is not only platform owners' business models that are important: Platform developers, integrators, managers, and users also need to be able to reach scale and scope 6G benefits to be successful. In pervasively profiting from 6G innovation, collaborative standards development, modularity, and the complementarity of technological solutions are important. This raises difficult openness, transparency and control, and collaboration vs. competition issues, especially in the use of data and algorithms such as AI, because the evolution of complementarities of different kinds is required to achieve the network effect. Technological complementarities are also required to ensure the various technological innovations created complement each other in commercialization. This is important if we wish 6G to become a pervasive general-purpose rather than merely an enabling technology with complex technical dependencies that are difficult to avoid when separate companies commercialize different 6G technologies and solutions (input oligopoly complementarity). Finally, consumer and production complementarities are required to efficiently regulate, standardize, and balance the supply and demand of 6G services.

However, it is not enough that technical, service, and business infrastructures will exist in the future 6G era. It is essential to consider whether users have real access to these services—that they have the required devices, and that they also know how to use them, as well as the available services. There is also a serious need to consider non-users and the reasons for their exclusion: Is it by their own choice or for another reason? A deeper understanding of technology in the form of design and development skills such as programming or digital fabrication also further enhances users' opportunities to play an active role in the ecosystem, and make and shape technologies for their personal needs. This also assists users in evaluating and reflecting on the technologies and their role in the user's own life, as well as more widely in society: Who benefits from technology or service use, and how? Who experiences value? What is the real price, and is it worth paying?

The limitation of this study is that the studied case, i.e. 6G, is in its birth phase as a business ecosystem (Moore 1993). Thus, the focal elements of ecosystemic business model scenarios will differ in phases further along in their lifecycle. Building on the discussion presented in this white paper, with roots in design, engineering, and economics research, there is a need for foresight research that explores the deployment of the ecosystemic business model in the 6G context, with a special focus on our three nested levels of experience, value creation, and sustainability. The major questions that could shape future research are as follows:

· Who owns the networks and has the incentives to invest in infrastructure?
· Who owns the radio spectrum and virtualized network resources, and how can these be utilized more dynamically in an affordable manner?
· Who owns the data and digital twins, how to get access to data and what added value can be generated using the data in communication networks?
· How can somatosensory experience of 6G technologies support new businesses models to derive value?
· How and why and what kind of platform-based ecosystemic business models can emerge in the future wireless systems context?
· What kind of successful open strategies and business models can emerge for 6G, and how? What role can a redefined sustainable economy or sharing economy principles play in future wireless?
· What kind of new business ecosystems could arise from 6G?

· How can new 6G ecosystem roles and business models emerge and change in different scenarios for those other than platform owners as well?
· What kind of dynamic capabilities will be needed in 6G to sense and seize opportunities?
· How can 6G become a truly general-purpose instead of an enabling technology?
· How can equal access to the offered services be ensured, rather than only for those living in highly developed countries?
· How can user empowerment be ensured in this context so users have a genuine opportunity to play an active role in the ecosystem, including design techniques for experimenting, making, and shaping technologies for their personal needs?
· How can users be assisted to take a reflective approach to technology use so they can evaluate and reflect on the role and value of technology in their everyday life contexts and more widely in society?
· How can novel regulation and governance principles such as community governance emerge in 6G?
· How can ubiquitous connected cognition bridge the physical, biological, and digital worlds?
· How can the emerging and required privacy and security regulation act as a trigger or barrier for new ecosystemic 6G business models?

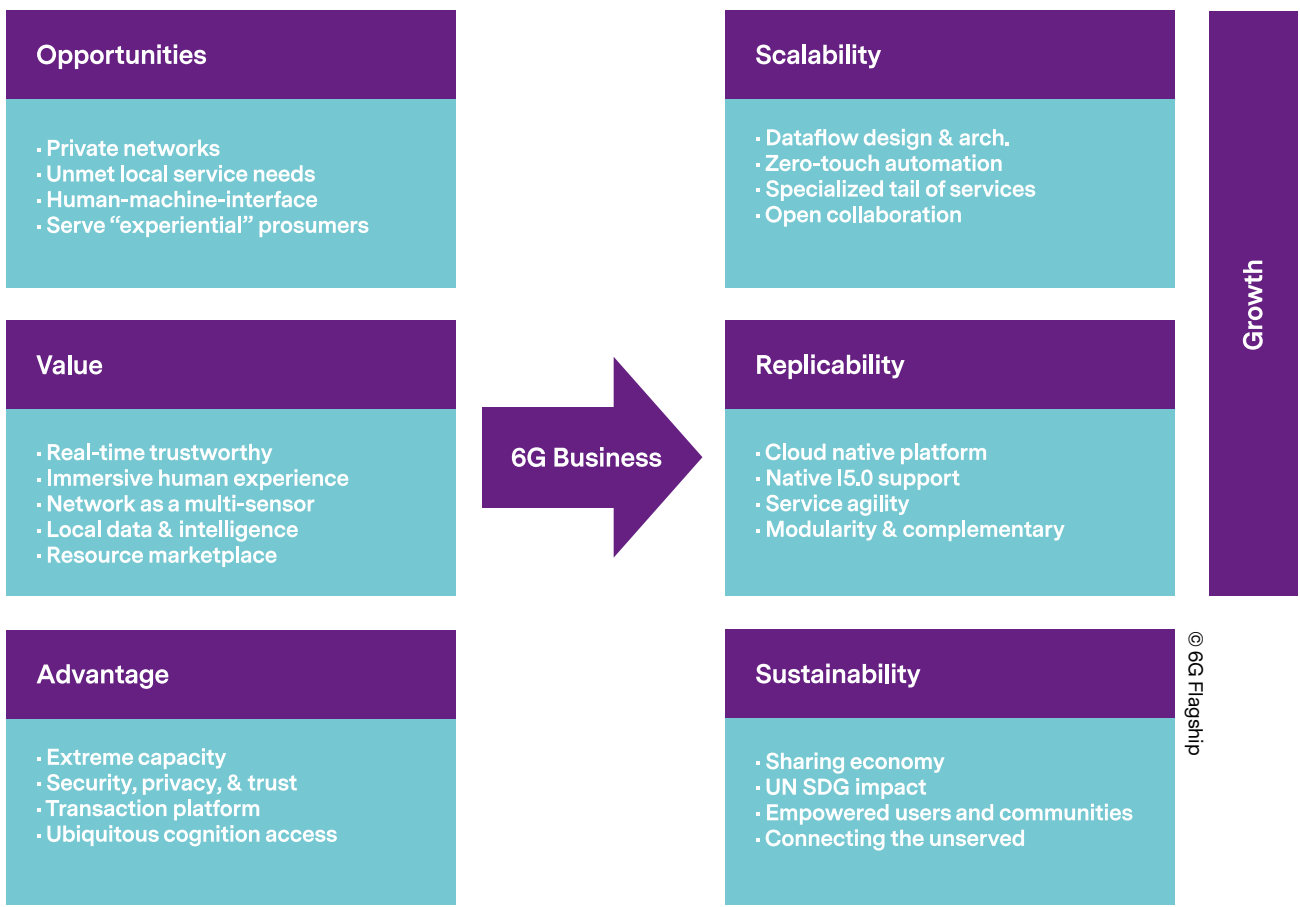

Figure 14. Summary of strategic 6G business model choices, activities, and their consequences

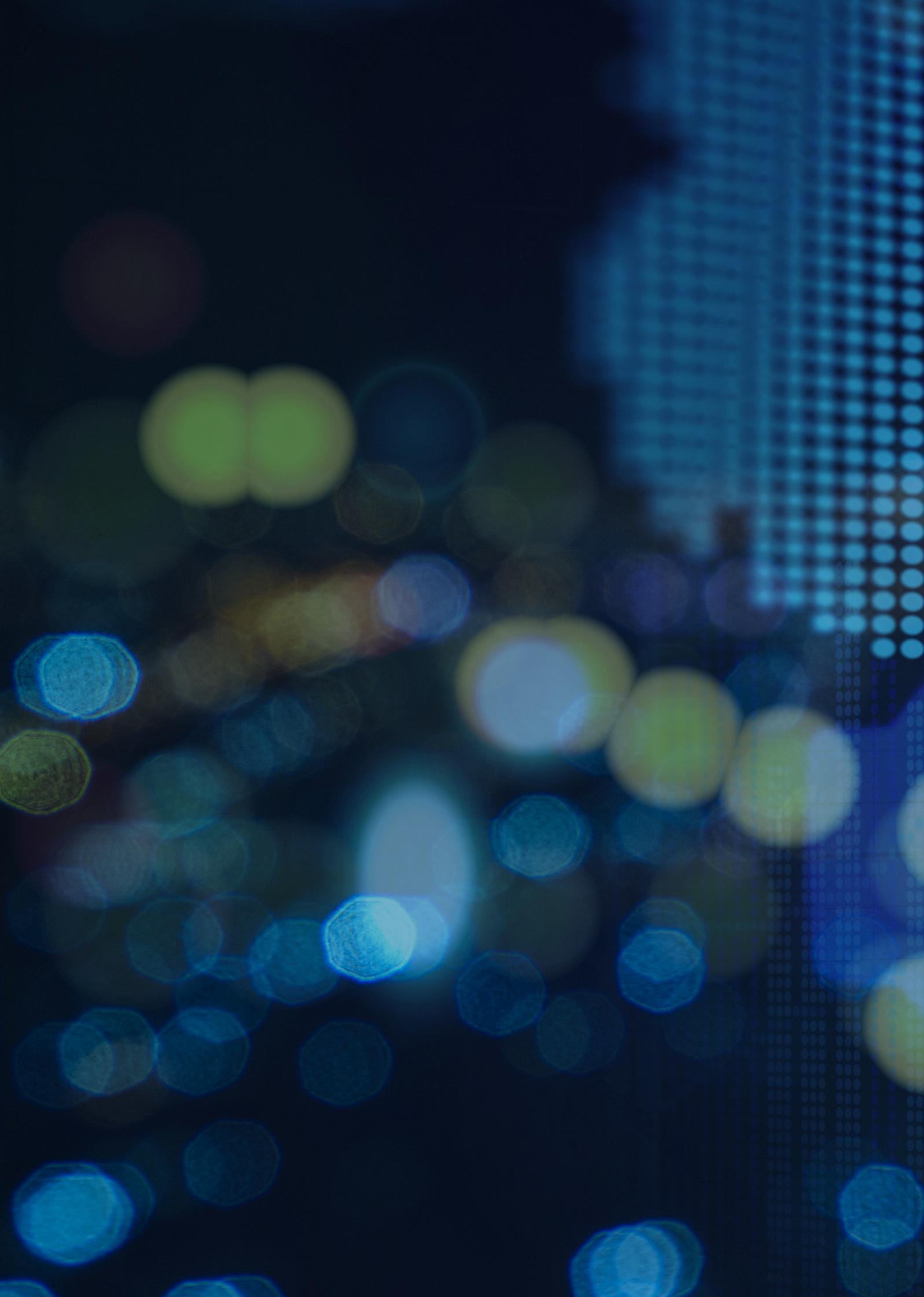

## White Paper on Business of 6G

**Editors:**
Seppo Yrjölä, University of Oulu, Nokia Enterprise, Oulu, Finland, seppo.yrjola@oulu.fi @nokia.com;
Petri Ahokangas, University of Oulu, Oulu Business School, Oulu, Finland; Marja Matinmikko-Blue,
University of Oulu, Centre for Wireless Communications, Oulu, Finland

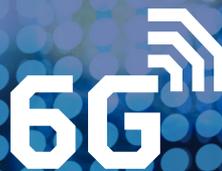

6gflagship.com